\documentclass[fleqn,usenatbib]{mnras}

\usepackage[T1]{fontenc}
\usepackage[utf8]{inputenc}
\usepackage{xcolor}

\DeclareRobustCommand{\VAN}[3]{#2}
\let\VANthebibliography\thebibliography
\def\thebibliography{\DeclareRobustCommand{\VAN}[3]{##3}\VANthebibliography}

\usepackage{acronym}
\usepackage{booktabs}
\usepackage{dcolumn}
\usepackage{makecell}
\usepackage{graphicx}
\usepackage{tabularx}
\usepackage{amsmath,amsfonts,amssymb}
\usepackage{mathrsfs}
\usepackage[normalem]{ulem}
\usepackage{multirow}
\usepackage{color}
\usepackage{natbib}
\usepackage{subcaption}

\usepackage{newtxtext,newtxmath}

\newcommand{\Msun}{\mathrm{\, M_\odot}}
\newcommand{\msun}{\mathrm{\, M_\odot}}

\newcommand{\up}[1]{^{\tiny{#1}}}
\newcommand{\low}[1]{_{\tiny{#1}}}
\newcommand{\err}[2]{\up{+ #2} \low{- #1}}
\def\Mo{{\rm M_{\odot}}}
\def\Mmax{M_\text{TOV}^\text{max}}

\def\md{M_{\rm ej}}
\def\vd{v_\infty}
\def\yd{Y_{e}}
\def\td{\theta_{\rm ej}}
\def\pd{\phi_{\rm ej}}
\def\sd{s_{\rm ej}}

\def\sdtd{\td^{\rm SD}}

\def\sdpd{\pd^{\rm SD}}
\def\mvd{\vd^{\rm med}}
\def\myd{\yd^{\rm med}}
\def\msd{\sd^{\rm med}}

\usepackage{pifont} 
\newcommand{\xmark}{\ding{55}}%

\DeclareOldFontCommand{\rm}{\normalfont\rmfamily}{\mathrm}
\DeclareOldFontCommand{\sf}{\normalfont\sffamily}{\mathsf}
\DeclareOldFontCommand{\tt}{\normalfont\ttfamily}{\mathtt}
\DeclareOldFontCommand{\bf}{\normalfont\bfseries}{\mathbf}
\DeclareOldFontCommand{\it}{\normalfont\itshape}{\mathit}
\DeclareOldFontCommand{\sl}{\normalfont\slshape}{\@nomath\sl}
\DeclareOldFontCommand{\sc}{\normalfont\scshape}{\@nomath\sc}
\DeclareRobustCommand*\cal{\@fontswitch\relax\mathcal}
\DeclareRobustCommand*\mit{\@fontswitch\relax\mathnormal}

\def\arraybackslash{\let\\\tabularnewline}

\definecolor{cyan}{rgb}{0,0.9,0.9}
\definecolor{orange}{rgb}{0.9,0.5,0}
\definecolor{magenta}{rgb}{1,0,1}
\definecolor{purple}{rgb}{0.8,0.4,0.8}
\definecolor{darkgreen}{rgb}{0.0,0.5,0.0}
\definecolor{gray}{rgb}{0.8242,0.8242,0.8242}
\definecolor{cadmiumgreen}{rgb}{0.0, 0.42, 0.24}
\definecolor{olive}{rgb}{0.5, 0.5, 0.0}

\newcommand{\Dudi}{Dudi \textit{et. al.}}

\newcommand{\reftab}[1]{{Table~\ref{#1}}}
\newcommand{\refsec}[1]{{Sec.~\ref{#1}}}
\newcommand{\reffig}[1]{{Fig.~\ref{#1}}}
\newcommand{\refapp}[1]{{Appendix~\ref{#1}}}
\newcommand{\refeq}[1]{{Eq.~(\ref{#1})}}

\newacro{BNS}{binary neutron star}
\newacro{NS}{neutron star}
\newacro{BH}{black hole}
\newacro{EOS}{equation of state}
\newacroplural{EOS}{equations of state}
\newacro{SD}{standard deviation}
\newacro{AMR}{adaptive mesh refinement}
\newacro{TOV}{Tolman–Oppenheimer–Volkoff}
\newacro{NSE}{nuclear statistical equilibrium}
\newacro{GW}{gravitational wave}
\newacro{GR}{general relativity}
\newacro{NR}{numerical relativity}
\newacro{ISCO}{innermost stable circular orbit}

\title[Numerical relativity simulations of GW190425]{Numerical relativity simulations of the neutron star merger GW190425: microphysics and mass ratio effects}

\author[A.~Camilletti et al.]{Alessandro Camilletti$^{1,2}$\thanks{Contact e-mail: a.camilletti@unitn.it},
	Leonardo Chiesa$^{1,2}$,
	Giacomo Ricigliano$^{3}$,
	Albino Perego$^{1,2}$,\newauthor
	Lukas Chris Lippold$^{4}$,
	Surendra Padamata$^{5,6}$,
	Sebastiano Bernuzzi$^{4}$,
	David Radice$^{5,6,7}$,\newauthor
	Domenico Logoteta$^{8,9}$
	and Federico Maria Guercilena$^{2,1}$
	\\
	$^{1}$ Dipartimento di Fisica, Universit\'{a} di Trento, Via Sommarive 14, 38123 Trento, Italy \\
	$^{2}$ INFN-TIFPA,Trento Institute for Fundamental Physics and Applications, via Sommarive 14, I-38123 Trento, Italy \\
	$^{3}$ Technische Universit{\"a}t Darmstadt, Institut f{\"u}r Kernphysik, Schlossgartenstr. 2, D-64289 Darmstadt, Germany \\
	$^{4}$ Theoretisch-Physikalisches Institut, Friedrich-SchillerUniversit{\"a}t Jena, 07743, Jena, Germany \\
	$^{5}$ Institute for Gravitation \& the Cosmos, The Pennsylvania State University, University Park PA 16802, USA \\
	$^{6}$ Department of Physics, The Pennsyvlania State University, University Park, PA 16802, USA \\
	$^{7}$ Department of Astronomy \& Astrophysics, The Pennsyvlania State University, University Park, PA 16802, USA \\
	$^{8}$ Dipartimento di Fisica, Universit\`{a} di Pisa, Largo B.  Pontecorvo, 3 I-56127 Pisa, Italy \\
	$^{9}$ INFN, Sezione di Pisa, Largo B. Pontecorvo, 3 I-56127 Pisa, Italy
}

\date{Accepted 2022 August 16. Received 2022 August 16; in original form 2022 April 8}

\pubyear{2022}

\begin{document}
	\label{firstpage}
	\pagerange{\pageref{firstpage}--\pageref{lastpage}}
	\maketitle
	
	\begin{abstract}
		GW190425 was the second gravitational wave (GW) signal compatible with a \ac{BNS} merger detected by the Advanced LIGO and Advanced Virgo detectors. Since no electromagnetic counterpart was identified, whether the associated kilonova was too dim or the localisation area too broad is still an open question.
		We simulate 28 \ac{BNS} mergers with the chirp mass of GW190425 and  mass ratio $1\leq q \leq 1.67$, using numerical-relativity simulations with finite-temperature, composition dependent \ac{EOS} and neutrino radiation.
		The energy emitted in GWs is $\lesssim 0.083\msun c^2$ with peak luminosity of $1.1$ - $2.4\times~10^{58}/(1+q)^2{\rm{erg~s^{-1}}}$. Dynamical ejecta and disc mass range between $5\times~10^{-6}$ - $10^{-3}$ and $10^{-5}$ - $0.1~\Msun$, respectively. Asymmetric mergers, especially with stiff \acp{EOS},  unbind more matter and form heavier discs compared to equal mass binaries. The angular momentum of the disc is $8$ - $10\Msun~GM_{\rm{disc}}/c$ over three orders of magnitude in $M_{\rm{disc}}$.
		While the nucleosynthesis shows no peculiarity, the simulated kilonovae are relatively dim compared with GW170817. For distances compatible with GW190425, AB magnitudes are always dimmer than $\sim20~{\rm{mag}}$ for the $B$, $r$ and $K$ bands, with brighter kilonovae associated to more asymmetric binaries and stiffer \acp{EOS}.
		We suggest that, even assuming a good coverage of GW190425's sky location, the kilonova could hardly have been detected by present wide-field surveys and no firm constraints on the binary parameters or \ac{EOS} can be argued from the lack of the detection.
	\end{abstract}
	
	

	
	\section{Introduction}
	
	The advent of the network of terrestrial gravitational wave (GW) detectors
	formed by Advanced LIGO \citep{TheLIGOScientific:2014jea} and Advanced Virgo
	\citep{TheVirgo:2014hva}, recently joined also by KAGRA
	\citep{Aso:2013eba,Akutsu:2018axf}, has opened the era of GW astronomy. At the
	end of the third observing run, the GW emission resulting from the late inspiral
	or from the merger of two \acp{BH}, one \ac{BH} and a \ac{NS}, or
	two \acp{NS} have all been observed
	\citep{LIGOScientific:2018mvr,LIGOScientific:2020ibl,LIGOScientific:2021djp}.
	
	So far, two GW signals compatible with the inspiral of a \ac{BNS} system were
	reported : GW170817 \citep{TheLIGOScientific:2017qsa} and GW190425
	\citep{Abbott:2020uma}. GW170817 was interpreted as the merger of a \ac{BNS} system
	with a chirp mass $\mathcal{M}_{\rm chirp}=(1.186 \pm 0.001) \Msun$. The masses
	of the individual stars were $M_A = (1.46^{+0.12}_{-0.10}) \Msun$ and $M_B =
	(1.27^{+0.09}_{-0.09}) \Msun$, at 90 per cent credible level, resulting in a
	total mass in the range $2.72-2.76 \Msun$
	\citep{TheLIGOScientific:2017qsa,Abbott:2018wiz}. The total mass of such a
	system is thus well within the expected range of Galactic \ac{BNS} systems, as
	resulting from electromagnetic (EM) observations of pulsars in \ac{BNS} systems
	\citep[see e.g.][]{Ozel:2016oaf}. The NS nature of the colliding objects was
	further corroborated by the detection of several EM counterparts originated from
	a galaxy located at 40Mpc from us, including a short gamma-ray burst and its
	afterglow, a kilonova, and possibly the non-thermal emission produced by the
	high speed tail of the dynamical ejecta expelled in the merger \citep[see
	e.g.][and references therein]{Radice:2020ids, Margutti:2020xbo}. The
	possibility of detecting GW170817 counterparts crucially depended on the
	availability of three detectors, which drastically reduced the sky localisation
	area to 16 deg$^2$
	\citep{TheLIGOScientific:2017qsa,GBM:2017lvd,LIGOScientific:2020ibl}.
	
	GW190425 represented a significantly different event with respect to GW170817 in
	many aspects \citep{Abbott:2020uma,LIGOScientific:2021djp}. The rest-frame chirp
	mass was $(1.44 \, \pm \, 0.02) \; \Mo$, while the NS mass ranges were $M_A =
	(2.0^{+0.6}_{-0.3}) \Msun$ and $M_B = (1.4^{+0.3}_{-0.3}) \Msun$, at 90 per cent
	credible level, resulting in a total mass in the range $3.3-3.7 \Msun$. Such a
	high total mass qualifies GW190425 as a possible outlier in the Galactic \ac{BNS}
	system distribution \citep{Abbott:2020uma,LIGOScientific:2021psn}.
	During the passage of the GW signal, the Livingston LIGO detector was offline
	and Virgo was unable to contribute to the measure because of the small
	signal-to-noise ratio (2.5) resulting from the large inferred distance ($D
	\approx 70-250$~Mpc). The effective presence of only one GW detector did not
	allow a good sky localisation ($\sim 10^4~{\rm deg}^2$).
	Despite an intense followup campaign within the first days after the \ac{GW}
	detection, no firm identification of EM counterparts was possible so far
	\citep[see e.g.][]{Coughlin:2019xfb,Steeghs.etal:2019}. In particular, the
	GROWTH and GRANDMA collaborations performed dedicated follow-up campaigns.
	GROWTH made use of the Zwicky Transient Facility (ZTF) and the Palomar
	Gattini-IR telescopes. The ZTF system covered 21 per cent of the probability
	integrated skymap and achieved a depth of 21 AB magnitudes in the $g$- and
	$r$-bands, while Palomar Gattini-IR covered 19 per cent of the probability
	integrated skymap in $J$-band to a depth of 15.5 mag \citep{Coughlin:2019xfb}.
	With 9 of its 21 heterogeneous telescopes, the GRANDMA network imaged 70
	galaxies covering $\lesssim 2$ per cent of the probability integrated skymap,
	attaining a depth of 17-23 AB magnitudes depending on the telescope
	\citep{Antier:2020nuy}. In absence of an optical or infrared counterpart,
	Apertif-WSRT searched for afterglow radio emission in a $9.5~{\rm deg}^2$ region
	of the high probability skymap \citep{Boersma:2021gyq}. Despite the reduced
	fraction of the covered skymap, the apparent lack of EM counterparts and the
	unusually high total mass of the binary leave open questions both on the origin
	of the system and on the remnant properties.
	
	Numerical modelling of \ac{BNS} mergers is a necessary step to properly
	interpret results, address open questions, and extract the largest amount of
	information from available data, even from the potential lack of detections. In
	particular, simulations of the inspiral, merger and early merger aftermath allow
	to extract the GW signal, the properties of the so-called dynamical ejecta, and
	the properties of the merger remnant \citep[see e.g.][for recent
	reviews]{Baiotti:2016qnr,Shibata:2019wef,Radice:2020ddv,Bernuzzi:2020tgt}.
	GW170817 was the privileged target of several simulation campaigns in numerical
	relativity \citep[see e.g.][]{Nedora:2020pak}. Recently, an independent study on
	GW190425 in numerical relativity has been proposed in \cite{Dudi:2021abi}
	(hereafter \Dudi). The authors set up 36 \ac{BNS} simulations targeted to
	GW190425 considering four mass ratios and three nuclear \acp{EOS} at different
	resolutions. They used cold \acp{EOS} with a density dependent composition fixed
	by neutrino-less beta-equilibrium conditions, and with thermal effects included
	by an effective $\Gamma$-law. \Dudi{} compute kilonova light curves employing a
	wavelength-dependent radiative transfer code \citep{Kawaguchi:2019nju}, for
	which the post-merger ejecta composition is fixed for all components. They
	concluded that, assuming an effective coverage of the event localisation region
	in the GROWTH follow-up campaign, the lack of kilonova detection suggests that
	GW190425 is incompatible with a face-on, unequal \ac{BNS} merger with more than 20
	per cent of mass difference between the two NSs. In all other cases (soft
	\acp{EOS}, edge-on and more distant mergers, or more symmetric binaries) the
	lack of detection is still compatible with a fainter kilonova signal.
	
	Several other works focused on GW190425 have recently appeared. For example \citet{Han:2020qmn} and \citet{Kyutoku:2020xka} investigated the possibility that GW190425 originated from a \ac{BH}-\ac{NS} merger by studying the corresponding \ac{GW} and kilonova signal, respectively.
	In \citet{Raaijmakers:2021slr} and \citet{Barbieri.etal:2021} kilonova light
	curves for GW190425 were computed under the assumption that the originating
	event was a \ac{BH}-\ac{NS} or a \ac{BNS} merger\footnote{In both works, the focus was broader
		than GW190425 kilonova characterisation, but this event was extensively studied
		as realistic test case.}. In both cases, the properties of the ejecta powering
	the kilonova signal were computed using fitting formulae derived from broad
	simulation samples, while the kilonova signals were computed using models with
	different levels of sophistication. In \citet{Barbieri.etal:2021}, the \ac{BNS}
	fitting formulae were taken from \cite{Radice:2018pdn} and from the appendix of
	\citet{Barbieri.etal:2021}. The \ac{NS} masses were chosen to be compatible with
	the GW190425 chirp mass, while the two employed \ac{NS} \acp{EOS} were
	compatible with present nuclear and astrophysical constraints. Additionally,
	using the same model, they also computed light curves directly using GW190425
	posteriors \citep{Abbott:2020uma}. They concluded that a light \ac{BH} in GW190425
	would have produced a brighter kilonova emission compared to \ac{BNS} case, allowing
	to distinguish the nature of the binary. However also in the \ac{BNS} case, the
	merger could have produced kilonovae bright enough to have been possibly
	detected by ZTF, especially for stiff \acp{EOS} and for more asymmetric systems.
	In \citet{Raaijmakers:2021slr}, only the posteriors from GW190425
	\citep{Abbott:2020uma} and the \ac{EOS} obtained from GW170817 analysis
	\citep{Abbott:2018exr} were used as input for the \ac{BNS} fitting formulae from
	\citet{Kruger:2020gig} and \citet{Foucart:2016vxd}. Based on the obtained ejecta
	and disc properties, kilonova light curves were computed using the semi-analytic
	model from \citet{Hotokezaka:2019uwo}. The latter adopts the radioactive heating
	rate fit from \citet{Korobkin:2012uy} and assumes a spherical symmetry for the
	ejecta geometry. Additionally, tests using the same kilonova model but fitting
	formulae from \citet{Radice:2018pdn,Barbieri.etal:2021,Dietrich:2020eud} were
	also performed.
	Despite these works, several open questions regarding GW190425 still remain. For
	example, how robust are the results obtained in numerical relativity for
	GW190425-like events? And, in particular, what is the impact of input physics
	that was so far neglected in GW190425-targeted simulations, including finite
	temperature, composition dependent \ac{EOS}, and neutrino radiation? What are
	the detailed properties of the dynamical ejecta expelled in these events and how
	do they depend on the binary properties and on the \ac{NS} \ac{EOS}? Is there a
	characteristic nucleosynthesis signature in these ejecta? Based on these
	results, what can we infer from the missing detection of electromagnetic
	counterparts for GW190425?
	
	To answer these questions, we setup 28 simulations in numerical relativity
	targeted to GW190425 with finite temperature, composition dependent \ac{NS}
	\acp{EOS}, and with neutrino radiation. We investigate the binary evolution up
	to the first $\approx 10$ ms after merger. We extract both remnant and dynamical
	ejecta properties, to give credible answers to some of the above questions. In
	particular, we use the detailed outcome of our simulations to compute
	nucleosynthesis yields and to set up kilonova models.
	We found that, for a distance compatible with GW190425, only in the case of a
	very stiff \ac{EOS} and a very asymmetric binary the resulting kilonova could
	have been bright enough to be observed by the ZTF facility. This suggests that
	the possible lack of kilonova counterpart for GW190425 provides much weaker
	constraints than previously thought.
	
	The paper is structured as follows: after a brief recap of the numerical setup
	and of the simulations properties in \refsec{sec:methods_and_models}, we resume
	the qualitative behaviour of the merger dynamics in \refsec{sec:merger_dynamics}
	and analyse the GW energetics in \refsec{sec:gravitational_waves}. The
	quantitative description of the remnant is reported in
	\refsec{sec:remnant_properties}, while we discuss the main properties of the
	dynamical ejecta in \refsec{sec:dynamical_ejecta}. In
	\refsec{sec:nucleosynthesis} and \refsec{sec:kilonovae} we describe the output
	from the nucleosynthesis process and its related kilonova signal. We compare our
	results with the one discussed in the literature in \refsec{sec:discussion}. We
	summarise our results in the conclusions in \refsec{sec:conclusions}.
	
	
	\begin{table*}
		\caption{NS initial properties grouped by \ac{EOS}. From left to right:
			\ac{EOS}, maximum \ac{TOV} mass $M_{\rm TOV}^{\rm max}$, maximum \ac{TOV} compactness
			$C_{\rm TOV}^{\rm max}$, NS masses $M_{\rm A}, M_{\rm B}$, total
			gravitational mass $M$, \ac{BNS} mass ratio $q \equiv M_{\rm A}/M_{\rm B}$,
			compactness of the two NSs $C_{\rm A}$, $C_{\rm B}$, tidal deformability of
			the \ac{BNS} $\tilde \Lambda$ defined in \refeq{eq:tidal_deformability}, the
			coefficient $k_2^{\rm L}$ defined in equation 4 of \citet{Zappa:2017xba},
			\refeq{eq:tidal_k2L}, the initial GW frequency $f_{\rm GW}(0)$, the total
			ADM mass of the system $M_{\rm ADM}$ and the initial ADM angular momentum
			$J_{\rm ADM}$.}
		\label{tab:sim}
		\begin{tabular}{ccc|cccccccc|ccc}
			\hline\hline
			EOS & $\Mmax$ & $C_\text{TOV}^\text{max}$ & $M_{\rm A}$ & $M_{\rm B}$ & $M$
			& $q$ & $C_{\rm A}$ & $C_{\rm B}$ & $\tilde{\Lambda}$ & $\kappa_2^{\rm L}$ & $f_\text{GW}(0)$ & $M_{\rm ADM}$ & $J_{\rm ADM}$\\
			& $[\Msun]$ & & $[\Msun]$ & $[\Msun]$
			& $[\Msun]$ & & & & & & $[\text{Hz}]$& $[\Msun]$  & $[\Msun^2]$\\
			\hline
			BLh & $2.103$ & $0.299$ & $1.654$ & $1.654$ & 3.308 & $1.0$ & 0.201 & 0.201 & $129.525$ & 194.3 & $608$ & $3.272$ & 10.23\\
			BLh & $2.103$ & $0.299$ & $1.750$ & $1.557$ & 3.307 & $1.12$ & 0.215 & 0.187& $133.008$ & 198.6 & $603$ & $3.271$ & 10.19\\
			BLh & $2.103$ & $0.299$ & $1.795$ & $1.527$ & 3.322 & $1.18$ & 0.222 & 0.183 & $131.172$ & 195.0 & $609$ & $3.284$ & 10.23\\
			BLh & $2.103$ & $0.299$ & $1.914$ & $1.437$ & 3.351 & $1.33$ & 0.242 & 0.172 & $134.612$ & 196.8 & $611$ & $3.313$ & 10.24\\
			\hline
			DD2 & $2.420$ & $0.300$ & $1.654$ & $1.654$ & 3.308 & $1.0$ & 0.184 & 0.184 & $257.963$ & 386.9 & $608$ & $3.270$ & 10.23\\
			DD2 & $2.420$ & $0.300$ & $1.795$ & $1.527$ & 3.322 & $1.18$ & 0.200 & 0.170 & $256.534$ & 382.8 & $609$ & $3.285$ & 10.24\\
			DD2 & $2.420$ & $0.300$ & $1.914$ & $1.437$ & 3.351 & $1.33$ & 0.214 & 0.160 & $254.057$ & 375.1 & $611$ & $3.312$ & 10.24\\
			DD2 & $2.420$ & $0.300$ & $2.149$ & $1.289$ & 3.438 & $1.67$ & 0.244 & 0.144 & $247.763$ & 354.8 & $616$ & $3.400$ & 10.25\\
			\hline
			SFHo & $2.059$ & $0.294$ & $1.654$ & $1.654$ & 3.308 & 1.0 & 0.209 & 0.209 & $101.708$ & 152.6 & $608$ & 3.275 & 10.25\\
			SFHo & $2.059$ & $0.294$ & $1.795$ & $1.527$ & 3.322 & $1.18$ & 0.230 & 0.191 & $102.689$ & 152.7 &$609$ & $3.290$ & 10.26\\
			SFHo & $2.059$ & $0.294$ & $1.914$ & $1.437$ & 3.351 & $1.33$ & 0.251 & 0.179 & $104.653$ & 153.0 & $611$ & $3.320$ & 10.28\\
			\hline
			SLy4 & $2.055$ & $0.303$ & $1.654$ & $1.654$ & 3.308 & $1.0$ & 0.212 & 0.212 & 89.251 & 133.9 & $608$ & $3.271$ & 10.23\\
			SLy4 & $2.055$ & $0.303$ & $1.795$ & $1.527$ & 3.322 & $1.18$ & 0.234 & 0.194 & 90.538 & 134.6 & $609$ & $3.285$ & 10.24\\
			SLy4 & $2.055$ & $0.303$ & $1.914$ & $1.437$ & 3.351 & $1.33$ & 0.256 & 0.181 & 93.140 & 136.0 & $611$ & $3.314$ & 10.25\\ 
			\hline\hline   
		\end{tabular}
	\end{table*}
	
	\section{Methods and Models}\label{sec:methods_and_models}
	
	\subsection{Binary merger calculations}
	\begin{figure*}
		\includegraphics{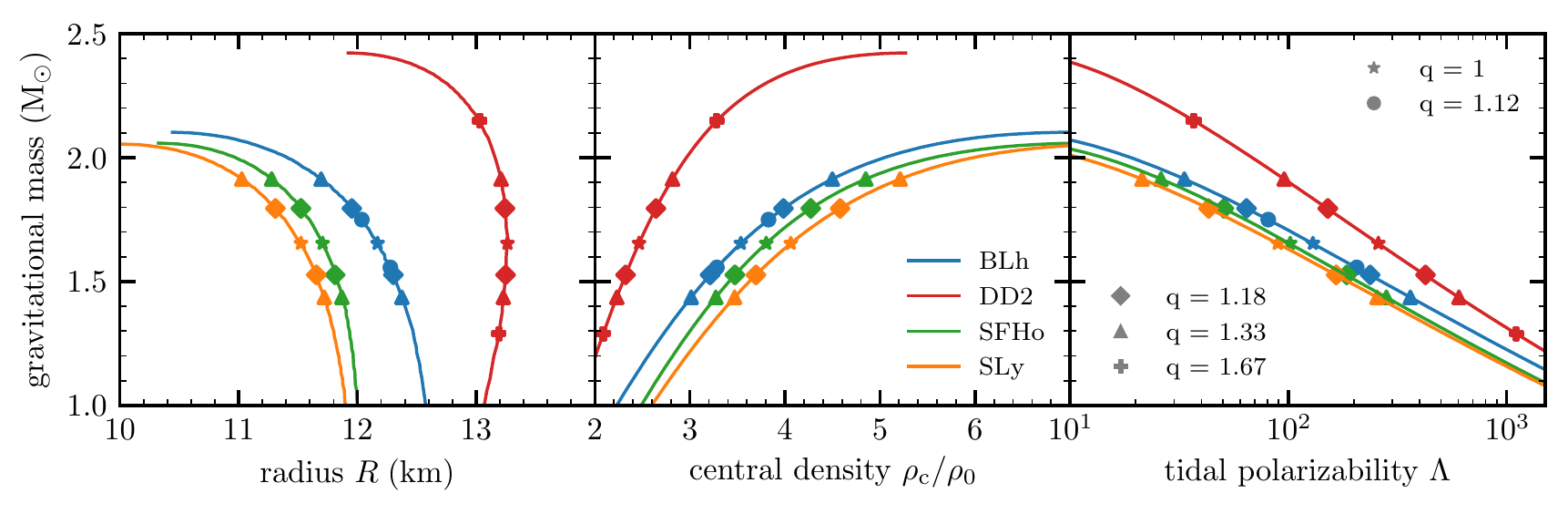}
		\caption{\ac{TOV} sequences for the \ac{NS} \acp{EOS} used in this work. Left
			panel: gravitational mass versus radius. Central panel: gravitational mass
			versus central density normalised to the nuclear saturation density,
			$\rho_0 = 2.67 \times 10^{14} \; {\rm g}~{\rm cm}^{-3}$. Right panel:
			gravitational mass versus tidal polarizability $\Lambda$. The different
			markers refer to the different mass ratios of the binaries evolved in the
			simulations.}
		\label{fig:TOV}
	\end{figure*}
	
	We evolve \ac{BNS} systems in full \ac{GR} through 3+1 numerical relativity simulations
	encompassing the latest orbits, the merger and the early post-merger phase.
	The spacetime metric is evolved with the Z4c formulation of Einstein's equations
	\citep{Bernuzzi:2009ex, Hilditch:2012fp} using the \texttt{CTGamma} code
	\citep{Pollney:2009yz, Reisswig:2013sqa}, developed within the
	\texttt{EinsteinToolkit} framework
	\citep{Loffler:2011ay,steven_r_brandt_2021_5770803}.
	We use the \texttt{WhiskyTHC} code \citep{Radice:2012cu, Radice:2013hxh},
	implemented within the \texttt{Cactus} \citep{Goodale:2002a, Schnetter:2007rb}
	framework to solve the \ac{GR} hydrodynamic equations. \texttt{WhiskyTHC}
	evolves the proton and neutron number density equations, in addition to the
	relativistic version of the momentum and energy conservation equations, written
	in conservative form.
	To properly resolve the NS structure and merger dynamics, and at the same time
	track the evolution of the ejecta on a large enough domain, we employ a mesh refinement \citep{Schnetter:2003rb,Reisswig:2012nc}
	consisting in seven nested grids characterised by a 1:2 linear scaling between
	consecutive grids, with the most refined level covering the two \acp{NS} during the inspiral and the central remnant after merger. We characterise each simulation by the resolution of the
	innermost grid, $h$, and in particular $h \approx 246 \; {\rm m}$ for low
	resolution (LR) and $h \approx 185 \; \rm m$ for standard resolution (SR) runs.
	Once the symmetry along the $z=0$ plane is taken into account, the simulated
	space is a cube of side $3024 \; \rm km$. For further details on the numerical
	setup we refer to \citet{Radice:2018pdn}.
	Thanks to the use of a puncture gauge, the spacetime evolution can handle the
	formation of a singularity within the computational domain
	\citep{Thierfelder:2010dv, Dietrich:2014wja}. The apparent horizon (AH) can
	possibly be detected by the \texttt{AHFinderDirect} thorn
	\citep{Thornburg:2003sf} of the \texttt{EinsteinToolkit}, from which the BH
	properties can be extracted.
	
	In all simulations we include compositional and energy changes due to the
	emission and absorption of neutrinos of all flavours. In particular, a grey
	leakage scheme \citep{Ruffert:1995fs, Neilsen:2014hha,Galeazzi:2013mia} is used
	to model the net neutrino emission rates both from optically thick regions,
	where neutrinos are expected to form a diffusing gas in thermal and weak
	equilibrium with matter, and optically thin regions. Neutrinos are then
	transported by an \texttt{M0} scheme \citep{Radice:2018pdn} through optically
	thin regions, where the reabsorption of streaming electron flavours
	(anti)neutrinos can happen in addition to local emission.
	
	We use four finite-temperature, composition-dependent \acp{EOS} compatible with
	current astrophysical \citep{Cromartie:2019kug,Miller:2019cac,Riley:2019yda} and
	nuclear \citep{Capano:2019eae, Jiang:2019rcw} constraints: BLh
	\citep{Bombaci:2018ksa, Logoteta:2020yxf}, HS(DD2) \citep{Typel:2009sy,
		Hempel:2009mc}, SFHo \citep{Steiner:2012rk} and SRO(SLy4)
	\citep{Douchin:2001sv,daSilvaSchneider:2017jpg}. In the following, we will refer
	to the second and fourth ones simply as DD2 and SLy4. All these \acp{EOS}
	include neutrons, protons, nuclei, electrons, positrons, and photons as relevant
	thermodynamics degrees of freedom, and assume baryon matter in nuclear
	statistical equilibrium.
	The BLh \ac{EOS} \cite{Logoteta:2020yxf} is an extension of the zero-temperature
	BL \ac{EOS} \cite{Bombaci:2018ksa} that includes finite-temperature effects and
	arbitrary particle composition. It was obtained within the finite-temperature
	version of the Brueckner–Bethe–Goldstone quantum many-body theory in the
	Brueckner–Hartree–Fock approximation. The underlying two-body and three-body
	interactions have been derived in chiral perturbation theory taking into account
	the effect of nucleon-nucleon and nucleon-nucleon-nucleon interactions.
	DD2 and SFHo were computed in the framework of relativistic mean field theories.
	The two \acp{EOS} differ because of the different parameterizations and coupling
	constants for modelling the mean-field nuclear interactions. The transition to
	inhomogeneous nuclear matter was done using an excluded volume approach.
	The SLy4 used in the present work is the finite temperature extension of the
	Skyrme effective nuclear interactions introduced in \citet{Douchin:2001sv}. The
	SLy4 \ac{EOS} reproduces well empirical saturation properties of nuclear matter
	as well as several observables deduced from the mass of finite nuclei.
	In \reffig{fig:TOV} we show the mass-radius, the mass-central density and the
	mass-quadrupolar tidal polarizability relations computed for the equilibrium NS
	sequences for the different \acp{EOS} used in this work. The quadrupolar tidal
	polarizability is computed as $\Lambda = (2/3) k_2 C^{-5}$,
	where $k_2$ is the dimensionless quadrupolar Love numbers \citep{Damour:1983a,
		Hinderer:2007mb}, and $C$ the stellar compactness $C = G M/ (c^2 R) $. The SLy
	\ac{EOS} produces the most compact \acp{NS}, while NSs modelled with the DD2
	\ac{EOS} have the largest radii around 13km for $1 \msun \lesssim M \lesssim 2.1
	\msun $.
	
	Initial data for our simulations are constructed using the pseudo spectral
	elliptic solver \texttt{Lorene} \citep{Gourgoulhon:2000nn}, using the \ac{EOS}
	slice at the lowest available temperature and assuming neutrino-less
	beta-equilibrium. All simulations are initialised as irrotational binaries on
	quasicirular orbits of coordinate radius $45 \; {\rm km}$. The residual initial eccentricity, 
	estimated following \citet{Kyutoku:2014yba}, is between 0.02 and 0.06 for all models.
	
	We set up and analyse a total of 28 simulations, 14 at SR and 14 at LR. In
	\reftab{tab:sim} we report a summary of all initial parameters characterising
	our simulations, in particular: the values of the individual stellar masses
	$M_{A,B}$ with $M_A > M_B$, the total gravitational mass $M$, the mass ratio $q
	\equiv M_A/M_B > 1$, the total ADM mass and angular momentum of the system
	$M_{\rm ADM}$ and $J_{\rm ADM}$, the stellar compactness $C_i$ for $i=A,B$, the
	the tidal deformability of the binary, $\tilde \Lambda$, defined as:
	\begin{equation}\label{eq:tidal_deformability}
	\tilde \Lambda = \frac{16}{13} \frac{(M_A + 12M_B) M_A^4}{M^5} \,
	\Lambda_A  + A \leftrightarrow B \, ,
	\end{equation}
	and the coefficients $k_2^{\rm L}$ as defined in equation 4 of
	\citet{Zappa:2017xba}, namely:
	\begin{equation} \label{eq:tidal_k2L}
	\kappa_2^L = 6 \left[ \frac{(3 M_B + M_A) M_A^4}{M^5}
	\Lambda_A + A \leftrightarrow B \right] \, ,
	\end{equation}
	where the notation $(A\leftrightarrow B)$ indicates a second term identical to
	the first except that the indices $A$ and $B$ are exchanged. We also report the GW initial frequency $f_{\rm GW}(0)$ measured in Hertz. All	\ac{BNS} parameters are compatible with the ones inferred from the GW signal GW190425
	\citep{Abbott:2020uma} using both the low- and high-spin priors, except for the
	ones characterised by $q = 1.33$ and $q = 1.67$, which are compatible only with
	high-spin prior.
	
	To better characterise the binaries used in this work and their properties in
	relation to the different \acp{EOS}, in \reffig{fig:TOV} we also highlight the
	properties of the NSs initially forming the binaries evolved by our simulations.
	Note that the initial conditions span a broad range of central densities, from
	$2.2 \rho_0$ to $6.0 \rho_0$ (in terms of the nuclear saturation density
	$\rho_0=2.67 \times 10^{14} \; {\rm g}~{\rm cm}^{-3}$) depending on the \ac{EOS}
	and mass ratio. For the more asymmetric binaries, the central density of the
	heaviest NS is roughly $1.5$ times larger than the one of the lightest \ac{NS},
	while in the equal mass case the two identical NSs have a central density $\sim
	1.2$ times larger than the one of the lightest NS in our sample.
	The single star tidal polarizability varies between two orders of magnitudes
	and, again, to asymmetric \ac{BNS} corresponds two NSs with rather different tidal
	polarizability: a more compact and less deformable NS along with a larger and
	more deformable one. Interestingly, $\tilde{\Lambda}$ varies only by a few
	percents within the same \ac{EOS}, while it changes by almost a factor of three
	between the SLy4 and the DD2 \ac{EOS}.

	\subsection{GWs and remnant properties}\label{subsec:GW and remnant method}
	
	We analyse the GW signal of the \ac{BNS} mergers as extracted at a coordinate radius
	of $\approx 591 \; \rm km$ from the \ac{BNS} centre of mass for all the simulations
	in the present work. We simulate the last 3 to 4 orbits before merger. The
	latter is defined as the moment in retarded time at which the amplitude of the
	$l=m=2$ mode of the GW waveform reaches its maximum. The short inspiral phase
	and the prompt collapse of the remnant to a BH do not permit to test in detail
	inspiral-merger-post-merger waveform models. Instead, we focus on the
	characterisation of the GW emission during the inspiral, merger and post-merger
	phases through integrated and peak quantities.
	In particular, we define the rescaled total energy radiated in GWs, $e_{\rm
		GW}^{\rm tot}$, and the rescaled angular momentum of the remnant, $j_{\rm
		rem}$, as:
	\begin{equation}
	e_{\rm GW}^{\rm tot} = \frac{(M - M_{\rm ADM})c^2 +
		E_{\rm GW}^{\rm rad}}{\nu M c^2} \, ,
	\end{equation}
	and
	\begin{equation}
	j_{\rm rem} = \frac{J_{\rm ADM} - J_{\rm GW}^{\rm rad}}{\nu G M^2/c} \, ,
	\end{equation}
	where $E_{\rm GW}^{\rm rad}$ and $J_{\rm GW}^{\rm rad}$ are the energy and
	angular momentum radiated in GWs during the whole simulation, and $\nu$ is the
	symmetric mass-ratio, $\nu = M_A M_B / M^2$.
	
	Our remnants are characterised by the presence of a central BH surrounded by an
	accretion disc. We extract the properties of both from our simulations.
	In particular, we define the disc as the portion of the remnant outside the
	apparent horizon whose rest mass density is smaller than $10^{13}\ {{\rm g}~{\rm
			cm}^{-3}}$, \citep[see e.g.][]{Shibata:2017xdx}. Moreover, we express the
	mass of the BH as
	\begin{equation}\label{eq:bh_mass}
	M_{\rm BH}^2 = M_{\rm irr}^2 + \left(\frac{c J_{\rm BH}}{2 G M_{\rm irr}} \right)^2 \, ,
	\end{equation}
	where $M_{\rm BH}$ and $J_{\rm BH}$ are the gravitational mass and spin of the
	BH, respectively, while $M_{\rm irr}$ is the irreducible BH mass:
	\begin{equation}
	M_{\rm irr} = \frac{c^2}{G} \sqrt{\frac{A_{\rm H}}{16 \pi}} \, ,
	\end{equation}
	with $A_{\rm H}$ the AH area. For a Kerr-BH, the irreducible mass is a
	non-decreasing quantity and it coincides with the gravitational mass for non
	rotating BHs. In analogy with the Kerr solution, we define the dimensionless
	spin parameter as $a_{\rm BH} \equiv (c J_{\rm BH}) / (G M_{\rm BH}^2)$.
	The AH finder is able to give an estimate of such quantities by locating the AH
	of the singularity, albeit it is not guaranteed that it does locate the AH with
	sufficient accuracy. This issue can clearly have an impact on the estimated BH
	properties. We compare the gravitational mass provided by the AH finder with the
	expected BH mass
	\begin{equation}\label{eq:exp_M_BH}
	M_{\rm BH}^{\text{exp}} =
	M_{\text{\rm ADM}} - M_{\rm disc} - E_{\rm GW}^{\rm rad}/c^2 \, ,
	\end{equation}
	where $E_{\rm GW}^{\rm rad}$ is the total energy radiated in GWs. In the above
	expression, we have neglected the ejecta mass and for the disc we have
	considered only the rest-mass energy. Similarly, for the spin parameter we
	compute the expected value as:
	\begin{equation}\label{eq:exp_a_BH}
	a_{\rm BH}^{\rm exp} = \frac{c J_{\rm BH}^{\rm exp}}{G \left( M_{\rm
			BH}^{\rm exp} \right)^2} = \frac{c (J_{{\rm ADM}} - J_{\rm GW}^{\rm rad}
		- J_{\rm disc})}{G \left( M_{\rm BH}^{\rm exp} \right)^2} \, ,
	\end{equation}
	where $J_{\rm GW}^{\rm rad}$ is the angular momentum radiated in GWs and $J_{\rm
		disc}$ is the angular momentum of the surrounding disc.

	\subsection{Ejecta and nucleosynthesis calculations}
	\label{subsec:nucleosynthesis calculations}
	
	From each simulation we consider the dynamical ejecta as the matter that becomes
	unbound within the end of the simulation on the basis of the geodesic criterion,
	i.e., when $|u_t| \geq c$, where $u_t$ is the time-component of the
	four-velocity. The properties of the ejecta are determined as matter crosses a
	spherical detector of coordinate radius $r_{\rm E} = 200 G\Msun/c^2 \approx
	294\ \textrm{km}$, discretised in $N_\theta=51$ polar and $N_\phi=93$ azimuthal
	uniform angular bins. For the unbound matter, the speed reached at infinity is
	computed as $\vd = c \sqrt{1 - (c/u_t)^2}$.
	
	The distribution of nuclei within the expanding ejecta is computed using the
	same approach and the same input data as the ones reported in
	\citet{Perego:2020evn}, that we briefly summarise in the following. We note that
	a similar approach was already used in \citet{Radice:2016dwd, Radice:2018pdn,
		Nedora:2020pak}, but with different input data. To obtain time-dependent yield
	abundances we employ \texttt{SkyNet} \citep{Lippuner:2017tyn}, a publicly available nuclear network which
	computes the nucleosynthesis depending on the evolution of a given Lagrangian
	fluid element. We evolve several trajectories with
	different initial parameters, with the aim of modelling the long-term expansion of
	the unbound matter measured in the simulations at the detector.
	All the trajectories start in
	\ac{NSE} from an initial temperature of $T_0=6.0$ GK. The corresponding initial
	density, $\rho_0 \equiv \rho(s,Y_e,T=6\, \textrm{GK})$, is determined by the
	\ac{NSE} solver implemented in \texttt{SkyNet} depending on the initial values
	of the electron fraction $Y_e$ and of the specific entropy $s$. The subsequent
	evolution of the density is set by the expansion time-scale $\tau$, first as an
	exponentially decaying phase and then as a homologous expansion:
	\begin{equation}
	\label{eq:rho_evolution}
	\rho(t) = \begin{cases}
	\makebox[2.0cm]{$\rho_0\,e^{-t/\tau}$\hfill}\textrm{if}\ t\leq3\tau \, ,\\
	\makebox[2.0cm]{$ \rho_0\bigg(\dfrac{3\tau}{et}\bigg)^3$\hfill}
	\textrm{if}\ t>3\tau \, .
	\end{cases}
	\end{equation}
	Parametric nucleosynthesis calculations are repeated for a set of fluid elements
	characterised by different values of $s$, $\tau$ and $Y_e$, ranging on a
	$26\times18\times25$ regular grid that spans the typical conditions
	characterising the ejecta in compact binary mergers, i.e., $1.5\leq
	s\ [k_B\ \text{baryon}^{-1}]\leq300$, $0.5\leq\tau\ \text{[ms]}\leq200$ and
	$0.01\leq Y_e\leq 0.48$, approximately logarithmic in the two former parameters
	while linear in the latter. To compute the nucleosynthetic yields in the ejecta
	we take the convolution of the output given by \texttt{SkyNet} with the
	distribution of the ejecta properties extracted from the numerical simulation at
	$r_{\rm E}$. While $s$ and $Y_e$ are directly extracted from the numerical
	simulation, $\tau$ is computed following the procedure described in
	\citet{Radice:2016dwd,Radice:2018pdn}.
	
	\subsection{Kilonova light curves calculations}
	\label{subsec:kilonova calculations}
	
	In order to compute kilonova light curves from the outcome of our simulations,
	we employ the multi-component anisotropic framework presented in
	\citet{Perego:2017wtu}.
	In this framework, axial symmetry and symmetry with
	respect to the \ac{BNS} orbital plane are assumed, while the polar angle $\theta$ is
	discretised in $N_{\theta} = 30$ angular bins equally spaced in $\cos{\theta}$.
	The kilonova emission is then computed in a ray-by-ray fashion by summing up the
	photon fluxes coming from each angular slice, properly projected along the line
	of sight of an observer located at a polar angle $\theta_{\rm view}$. Inside
	each slice, a 1D kilonova model is used. The latter depends on the mass and
	(root mean square) speed of the ejecta, as well as on an effective grey opacity
	$\kappa$. Inside each ray, several ejecta components are considered, resulting
	from the expulsion of matter operated by different mechanisms, acting on
	different time-scales and providing distinct ejecta properties. The total
	luminosity is found by summing over the contributions of the different ejecta
	components, assuming that the energy emitted by the innermost ones is quickly
	reprocessed and emitted by the outermost component\footnote{The location of the
		components is determined by the location of the photospheres.}.
	
	Differently from the model originally implemented in \citet{Perego:2017wtu} and
	later employed, for example, in \citet{Radice:2018pdn, Radice:2018xqa,
		Breschi:2021tbm, Barbieri:2019kli, Barbieri:2019sjc, Barbieri.etal:2021}, here
	we adopt a new semi-analytical 1D kilonova model for each angular slice that we
	present in the following. The model assumes a spherically symmetric and
	optically thick outflow with a constant average grey opacity. The outflow
	expands with an homologous expansion law, i.e., the density of each fluid
	element decreases as $t^{-3}$ while its expansion speed stays constant, starting
	from a few hours after merger. The kilonova emission is calculated as the
	combination of two contributions, one emitted at the photosphere and one coming
	from the optically thin layers above it.
	The contribution coming from the photosphere is computed starting from the
	semi-analytic formula for the luminosity originally proposed by \citet{Wollaeger:2017ahm}
	and derived from a solution of the radiative transfer equation in the diffusion
	approximation \citep{Pinto:1999ai}.
	This formula was further validated in
	\citet{Wu:2021ibi}, where it showed a very reasonable agreement with results provided by the radiation hydrodynamics code \texttt{SNEC}.
	While the original model assumes that the whole ejecta are in optically
	thick conditions, an increasing fraction of it resides outside of the
	photosphere, becoming optically thin to thermal radiation.
	For this reason, the outcome of this computation is rescaled by a factor
	$M_{\mathrm{thick}}/M_{\mathrm{ej}}$, where $M_{\mathrm{thick}}$ is the mass of
	the optically thick part of the ejecta, defined as the region enclosed by the
	photosphere. The photospheric radius $R_{\mathrm{ph}}(t)$ is found analytically
	by imposing the condition $\tau_{\gamma}(R_{\mathrm{ph}})=2/3$, where
	$\tau_{\gamma}$ is the optical depth of the material, and by using the
	homologous density profile as in \citet{Wollaeger:2017ahm}:
	\begin{equation}\label{eq:density profile}
	\rho(t,x)=\rho_0\left(\frac{t_0}{t}\right)^3\left(1-x^2\right)^3,
	\end{equation}
	where $\rho_0$ is the density at the initial time $t_0$ and $x=v/v_{\rm max}$ is
	the dimensionless radial variable. The photospheric temperature
	$T_{\mathrm{ph}}(t)$ is computed from the photospheric luminosity and radius
	using the Stefan-Boltzmann law. A temperature floor of $2000$ $K$ for
	$T_{\mathrm{ph}}(t)$ is applied in order to account for electron-ion
	recombination in the expanding ejecta. When $T_{\mathrm{ph}}(t)$ reaches the
	temperature floor, $R_{\mathrm{ph}}(t)$ is redefined using again the
	Stefan-Boltzmann law. Furthermore a Planckian black body spectrum is assumed at
	the photosphere.
	
	The contribution to the luminosity from the thin part of the ejecta is computed
	by partitioning the latter into equal mass shells and by assuming that each
	shell with temperature $T$ emits its radioactive decay energy assuming local
	thermodynamics equilibrium. To characterise the temperature of the thin part of
	the ejecta, we adopt a temperature profile similar to the one derived in
	\citet{Wollaeger:2017ahm} under the assumption of radiation dominated,
	homologous expansion: $T(t,x)=T_0(x)\left(t_{\rm tr}(x)/t \right)$, where
	$T_0(x)$ is the temperature of the photosphere as it transits through the shell
	centred in $x$ at the time $t_{\rm tr}(x)$. The bolometric luminosity
	contribution from the thin region is computed by multiplying the mass of each
	shell by the specific heating rate.
	
	For the nuclear heating rates powering the kilonova emission, we employ the
	analytic fitting formula first presented in \citet{Wu:2021ibi} and based on the
	results from the nucleosynthesis calculations reported in
	\citet{Perego:2020evn}: $\dot{\epsilon}_{\mathrm{r}}(t)=At^{-\alpha}$, where $A$
	and $\alpha$ are fit parameters. The latter are interpolated from tabulated
	values on the same $(Y_e,s,\tau)$ grid used for the nucleosynthesis calculations
	(see \refsec{subsec:nucleosynthesis calculations}).
	A constant thermalisation efficiency $\epsilon_{\mathrm{th}}=0.5$ is employed
	for the thick region of the ejecta, while we construct a thermalisation
	efficiency profile for the thin part starting from the analytic fitting formula
	proposed in \citet{Barnes:2016umi}. The expression for the thermalisation
	efficiency profile reads:
	\begin{equation}\label{eq:thermalisation}
	\epsilon_{\mathrm{th}}(t,x)=
	0.36\left[\exp(-aX)+\frac{\ln(1+2bX^d)}{2bX^d}\right] \, ,\\
	\end{equation}
	where $a$, $b$ and $d$ are the fitting parameters reported in
	\citet{Barnes:2016umi} and interpolated from tabulated values on a grid spanning
	the intervals $1\times10^{-3} M_{\odot} < M_{\mathrm{ej}} < 5\times10^{-2}
	M_{\odot}$ and $0.1c < v_{\mathrm{ej}} < 0.3c$. In the original formulation of
	\citet{Barnes:2016umi}, obtained assuming $\rho(t) = \rho_0 (t/t_0)^3$,
	$X(t,x)=t$. Due to the use in our model of the density profile
	\refeq{eq:density profile}, we adopt $X(t,x)=t/(1-x^2)$, instead. In this work,
	we consider two ejecta components: a dynamical ejecta and a disc ejecta
	component, both symmetric with respect to the equatorial plane and to the polar
	axis. Following the same procedure described in \refsec{subsec:nucleosynthesis
		calculations}, we directly extract from the simulations the profiles of the
	properties of the dynamical component, namely the distributions of the ejecta
	mass, of the root mean square velocity at infinity, of the average electron
	fraction, average entropy and average density at the extraction radius, as a
	function of the polar angle $\theta$, averaged over the azimuthal angle $\phi$.
	The opacity $\kappa$ is computed by interpolating the results of the atomic
	calculations performed in \citet{Tanaka:2019iqp} for a wide range of the
	electron fraction $0.01\leq Y_e\leq0.50$. Additionally, inspired by disc simulations
	of \citet{Wu:2016pnw}, \citet{Lippuner:2017bfm}, \citet{Siegel:2017nub}, \citet{Fernandez:2018kax}, \citet{Fahlman:2022jkh}, we assume
	that a fraction between $\sim20$ and $\sim40$ per cent of the disc mass inferred
	from our simulations (see \refsec{sec:remnant_properties}) is ejected in the form
	of a viscosity-driven wind. We model the mass of this disc wind as uniformly
	distributed in $\theta$, as we do not expect preferential latitudes for the ejection.
	Moreover, for the disc ejecta we assume a root mean square velocity of $0.06c$,
	a uniform opacity of $5$ ${\rm cm^2~g^{-1}}$, an average entropy of $20$ $k_{\rm B}~{\rm
		baryon^{-1}}$ and an expansion time-scale of $30$ ms. We stress that our kilonova model relies on a large number of assumptions and simplifications which limit its accuracy. However, for the parameters that are not directly fixed by our simulations, we chose representative
	values in broad agreement with what obtained by fitting AT2017gfo data with the
	original kilonova model \citep{Perego:2017wtu}.
	
	
	\section{Results}
	\label{sec:results}
	

	\subsection{Merger Dynamics}
	\label{sec:merger_dynamics}
	
	All simulations in our sample follow a qualitative common evolution pattern with
	quantitative differences, mainly due to the different tidal deformability
	provided by the \acp{EOS} and \ac{BNS} mass ratios. All simulations result in the
	prompt collapse of the central part of the remnant into a BH. In this context,
	we say that a \ac{BNS} simulation has resulted in a prompt collapse if the minimum of
	the lapse function inside the computational domain decreases monotonically
	immediately after merger without showing core bounces.
	We define the moment of formation of the BH as the time at which the lapse
	function drops below $0.2$. In all simulations presented here the BH forms
	within a fraction of a ms after the merger ($t_{\rm BH} < 0.47~{\rm ms}$, see
	\reftab{tab:rem_props}).
	
	Tidal forces deform the NSs during the inspiral, especially the lighter and less
	compact one. This effect is more pronounced for \ac{BNS} with stiffer \acp{EOS},
	providing, for the same gravitational mass, a less compact NS. The subsequent
	merger dynamics is able to unbind matter from the tidal tails on a few dynamical
	time-scales. The neutron-rich matter ballistically expelled during this phase
	from the tidal tails has low entropy and can have large enough velocity to
	escape the potential barrier, contributing to the dynamical ejecta. The
	otherwise gravitationally bound matter forms a disc with toroidal shape around
	the forming BH. \ac{BNS} models characterised by a stiffer \ac{EOS} expel more matter,
	such that more dynamical ejecta and larger discs are found, as discussed in
	detail below.
	
	\begin{figure}
		\includegraphics[width=\columnwidth]{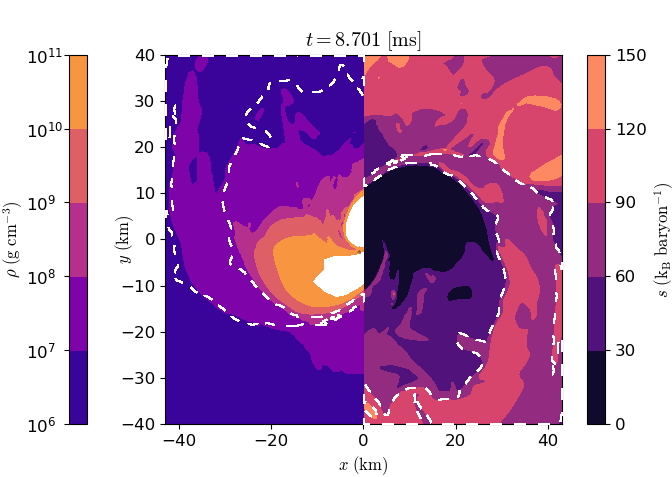}
		\caption{Snapshot of the rest mass density (left) and the entropy per baryon
			(right) taken at $\sim 0.3~{\rm ms}$ after BH formation across the orbital
			plane for the equal mass \ac{BNS} merger SR simulation with the SFHo \ac{EOS}.
			Matter inside the dashed contour with entropy $90-120~{\rm k}_{\rm B}~{\rm
				baryon}^{-1}$ and densities $< 10^8~{\rm g}~{\rm cm}^{-3}$ comes from
			the rotationally non-symmetric central object, expelled from the contact
			surface of the two stars. Since equal mass binaries eject few
			$10^{-5}~\Msun$, this shocked matter have a prominent role in the median
			properties of the ejecta.}
		\label{fig:dens_entr}
	\end{figure}
	During the few fractions of ms that precede \ac{BH} formation, a small amount of very high-entropy matter coming from the \ac{NS} contact interface is expelled, see \reffig{fig:dens_entr}. This extremely shocked matter is characterised by higher entropy and electron fraction than the ones that characterise matter expelled by tidal forces. This small component with entropy of $90 - 120~\rm{k_B}~\rm{baryon^{-1}}$ is responsible of the bimodal distribution of the entropy shown in \reffig{fig:ejecta_histograms}.
	Its unbound component contributes to the dynamical ejecta, while the bound mass contributes to the disc formation, spanning in both cases a broader polar angle than the bound and unbound matter of tidal origin.
	The resulting disc, ejecta and the central BH will be the focus of
	\refsec{sec:remnant_properties} and \refsec{sec:dynamical_ejecta}.
	
	\subsection{Gravitational-Wave Luminosity}
	\label{sec:gravitational_waves}
	
	\begin{table*}
		\caption{For each simulation the table reports the rescaled angular momentum
			of the remnant, $j_{\rm rem}$; the rescaled total energy radiated in GWs,
			$e_{\rm GW}^{\rm tot}$; the BH expected mass (spin), $M_{\rm BH}^{\rm
				exp}$ ($a_{\rm BH}^{\rm exp}$) as defined in \refeq{eq:exp_M_BH}
			(\refeq{eq:exp_a_BH}); the BH mass (spin) as detected from the AH finder,
			$M_{\rm BH}$ ($a_{\rm BH}$), together with the related average on a sample
			time, $\langle M_{\rm HB} \rangle$ ($\langle a_{\rm BH} \rangle$). We
			report values from the SR simulations and the error inside brackets
			estimated as the absolute semi-difference between the SR and LR values.
			Uncertainties refers to the least significant digit(s).}
		\label{tab:rem_props}
		\begin{tabular}{ccc|cccccccccccc}
			\hline\hline
			EOS & $q$ & AH finder
			& $t_{\rm BH}-t_{\rm mrg}$ & $j_{\rm rem}$ & $e_{\rm GW}^{\rm tot}$ & $L_{\rm peak}$ & $M_{\rm BH}^{\rm exp}$ & $M_{\rm BH}$ & $\langle M_{\rm BH} \rangle$ & $a_{\rm BH}^{\rm exp}$ & $a_{\rm BH}$ & $\langle a_{\rm BH} \rangle$
			\\ 
			& & & (ms) & & & $10^{55}~[{\rm erg}~{\rm s}^{-1}]$  &$[M_{\odot}]$ & $[M_{\odot}]$ & $[M_{\odot}]$ & & & \\
			\hline
			BLh & 1.0 & \checkmark
			&\makecell{0.185 \\ (2)} 
			&\makecell{2.994 \\ (8)} &\makecell{0.099 \\ (1)} &\makecell{ 8.23\\ (13)} 
			&\makecell{3.2259 \\ (2)}  &\makecell{3.2349 \\ (2)} &\makecell{3.245 \\ (2)} &\makecell{0.788 \\ (2)} &\makecell{0.7860 \\ (1)} &\makecell{0.801 \\ (2)} \\ 
			BLh & 1.12 & \checkmark
			&\makecell{0.209 \\ (2)}
			&\makecell{3.012 \\ (8)} &\makecell{0.097 \\ (1)} &\makecell{7.75 \\ (22)}
			&\makecell{3.2250 \\ (5)}  &\makecell{3.2330 \\ ($<10^{-1}$)} &\makecell{3.245 \\ (2)} &\makecell{0.789 \\ (2)} &\makecell{0.7865 \\ (3)} &\makecell{0.802 \\ (2)} \\ 
			BLh & 1.18 & \checkmark
			&\makecell{0.209 \\ (30)} 
			&\makecell{3.020 \\ (6)} &\makecell{0.098 \\ (1)} &\makecell{7.19 \\ (9)}
			&\makecell{3.2411 \\ (18)}  &\makecell{3.2458 \\ (4)} &\makecell{3.259 \\ (2)} &\makecell{0.789 \\ (2)} &\makecell{0.7866 \\ (1)} &\makecell{0.803 \\ (3)} \\
			BLh & 1.33  & \checkmark
			&\makecell{0.221 \\ (8)}  
			&\makecell{3.067 \\ (6)} &\makecell{0.090 \\ (1)} &\makecell{5.53 \\ (8)}
			&\makecell{3.2559 \\ (2)}  &\makecell{3.2573 \\ (6)} &\makecell{3.273 \\ (1)} &\makecell{0.780 \\ (5)} &\makecell{0.7779 \\ ($< 10^{-1}$)} &\makecell{0.796 \\ (3)} \\      
			\hline
			DD2 & 1.0 & \xmark
			&\makecell{0.422 \\ (10)}  
			&\makecell{3.122 \\ (9)} &\makecell{0.092 \\ (2)} &\makecell{5.46 \\ (18)}
			&\makecell{3.2210 \\ ~}  &\makecell{- \\ ~} &\makecell{- \\ ~} &\makecell{0.826 \\ ~} &\makecell{- \\ ~} &\makecell{- \\ ~} \\ 
			DD2 & 1.18 & \xmark
			&\makecell{0.445 \\ (6)}  
			&\makecell{3.117 \\ (6)} &\makecell{0.091 \\ (1)} &\makecell{4.96 \\ (12)}
			&\makecell{3.2298 \\ ~}  &\makecell{- \\ ~} &\makecell{- \\ ~} &\makecell{0.820 \\ ~} &\makecell{- \\ ~} &\makecell{- \\ ~} \\ 
			DD2 & 1.33 & \xmark
			&\makecell{0.469 \\ (41)} 
			&\makecell{3.149 \\ (2)} &\makecell{0.0877 \\ (2)} &\makecell{4.06 \\ (3)}
			&\makecell{3.2315 \\ ~}  &\makecell{- \\ ~} &\makecell{- \\ ~} &\makecell{0.780 \\ ~} &\makecell{- \\ ~} &\makecell{- \\ ~} \\
			DD2 & 1.67 & \xmark
			&\makecell{0.374 \\ (2)}  
			&\makecell{3.204 \\ (3)} &\makecell{0.077 \\ (3)} &\makecell{2.89 \\ (4)}
			&\makecell{- \\ }  &\makecell{- \\ } &\makecell{- \\ } &\makecell{- \\ } &\makecell{- \\ } &\makecell{- \\ } \\      
			\hline
			SFHo & 1.0 & \checkmark
			&\makecell{0.138 \\ (2)} 
			&\makecell{2.953 \\ (14)} &\makecell{0.102 \\ (1)} &\makecell{9.98 \\ (22)}
			&\makecell{3.223 \\ (1)}  &\makecell{3.25 \\ ~} &\makecell{3.26 \\ ~} &\makecell{0.778 \\ (1)} &\makecell{0.774 \\ ~} &\makecell{0.79 \\ ~} \\
			SFHo & 1.18 & \checkmark
			&\makecell{0.138 \\ (18)}  
			&\makecell{2.976 \\ (8)} &\makecell{0.097 \\ (1)} &\makecell{8.86 \\ (17)}
			&\makecell{3.240 \\ (1)}  &\makecell{3.27 \\ ~} &\makecell{3.28 \\ ~} &\makecell{0.776 \\ (2)} &\makecell{0.775 \\ ~} &\makecell{0.79 \\ ~} \\
			SFHo & 1.33 & \checkmark
			&\makecell{0.126 \\ (8)} 
			&\makecell{3.066 \\ (17)} &\makecell{0.0872 \\ (4)} &\makecell{7.32 \\ (16)}
			&\makecell{3.268 \\ ~}  &\makecell{3.29 \\ ~} &\makecell{3.29 \\ ~} &\makecell{0.783 \\ ~} &\makecell{0.770 \\ ~} &\makecell{0.79 \\ ~} \\
			\hline
			SLy4 & 1.0 & \xmark
			&\makecell{0.138 \\ (18)}  
			&\makecell{3.031 \\ (6)} &\makecell{0.105 \\ (1)} &\makecell{10.90 \\ (32)}
			&\makecell{3.2167 \\ (1)}  &\makecell{- \\ ~} &\makecell{- \\ ~} &\makecell{0.801 \\ (2)} &\makecell{- \\ ~ } &\makecell{- \\ ~} \\
			SLy4 & 1.18 & \xmark
			&\makecell{0.114 \\ (14)} 
			&\makecell{3.010 \\ (12)} &\makecell{0.103 \\ (1)} &\makecell{9.67 \\ (23)}
			&\makecell{3.2323 \\ (6)}  &\makecell{- \\ ~} &\makecell{- \\ ~} &\makecell{0.791 \\ (3)} &\makecell{- \\ ~} &\makecell{- \\ ~} \\
			SLy4 & 1.33 & \xmark
			&\makecell{0.114 \\ (2)} 
			&\makecell{3.043 \\ (9)} &\makecell{0.097 \\ (1)} &\makecell{7.97 \\ (7)}
			&\makecell{- \\ ~}  &\makecell{- \\ ~} &\makecell{- \\ ~} &\makecell{- \\ ~} &\makecell{- \\ ~} &\makecell{- \\ ~} \\
			\hline
			\hline
		\end{tabular}
	\end{table*}
	
	In the left columns of \reftab{tab:rem_props}, we report GW data (i.e., $j_{\rm
		rem}$, $e_{\rm GW}^{\rm tot}$, and $L_{\rm peak}$ ) as extracted from our
	GW190425-like \ac{BNS} simulations.
	We first test the quasi-universal relation between $e_{\rm GW}^{\rm tot}$ and
	$j_{\rm rem}$ given in \citet{Zappa:2017xba}: $e_{\rm fit}^{\rm tot}(j_{\rm
		rem}) = c_2 j_{\rm rem}^2 + c_1 j_{\rm rem} + c_0$, with $c_0 = 0.95$, $c_1 =
	-0.44$ and $c_2 = 0.053$ \footnote{We notice that, despite referring to the same
		fit, the fitting values reported in this work have one more figure than the ones
		originally reported by \citet{Zappa:2017xba}.}. These coefficients were fitted
	over a dataset containing more than 200 \ac{BNS} merger simulations performed with
	the \texttt{BAM} \citep{Brugmann:2008zz} and \texttt{THC} codes. The \ac{BNS}
	simulations were grouped in four categories according to the fate of the
	remnant: prompt collapse to a BH, short-lived hypermassive \ac{NS}, supramassive
	\ac{NS} and stable NS. This simple quadratic polynomial in $j_{\rm rem}$ was
	very effective in relating the angular momentum of the remnant with the total
	radiated energy in the whole dataset, despite the different fates of the
	remnants, nuclear \acp{EOS}, and intrinsic \ac{BNS} parameters.
	Moreover, the ranges $j_{\rm rem} \in [2.944, 3.204]$ and $e_{\rm GW}^{\rm tot}
	\in [0.077, 0.105]$ are compatible with the respective ranges presented in
	\citet{Zappa:2017xba} for the case of \ac{BNS} resulting in a prompt collapse. We
	notice that the absolute value of the relative error $\left| e_{\rm fit}^{\rm
		tot}- e_{\rm GW}^{\rm tot} \right| / e_{\rm GW}^{\rm tot}$ $\lesssim
	\mathcal{O}(0.1)$ is in accordance with the residuals plotted in figure~4 of
	\citet{Zappa:2017xba}. Additionally, we remark that $e_{\rm GW}^{\rm tot} <
	e_{\rm GW}^{\rm fit}$, also in accordance with the behaviour of the
	prompt-collapse simulations in \citet{Zappa:2017xba}. To further test the
	quality of the fit results with respect to the uncertainties of numerical origin
	we compute the ratio between the residuals and the estimated total error due to
	resolution uncertainties, $\sqrt{ {\delta e_{\rm GW}^{\rm tot}}^2 + \delta
		{e_{\rm fit}^{\rm tot}}^2 }$, where $\delta e_{\rm fit}^{\rm tot} = \sqrt{ 4
		c_2^2 j_{\rm rem}^2 + c_1^2} ~ \delta j_{\rm rem}$. The uncertainties of
	numerical origin, $\delta j_{\rm rem}$ and $\delta e_{\rm GW}^{\rm tot}$, are
	computed as the absolute value of the semi-difference between SR and LR results.
	The typical values are $\lesssim 1$, indicating that the numerical error
	accounts for a significant fraction of the observed discrepancy.
	Finally we emphasise that the rescaled GW peak luminosity, $(q/\nu)^2 \, L_{\rm
		peak}$, and $\kappa_2^L$ coefficient span the same range of the prompt
	collapse \acp{BNS} reported in figure 2 of \citet{Zappa:2017xba}, i.e., $[1.11,2.36]
	\times 10^{58}~{\rm erg}~{\rm s}^{-1}$ and $[134, 387]$, respectively. We recall
	that $\kappa_2^L$ is the coefficient that parametrises the leading effect of
	tides on the GW emission from a BNS merger in the post-Newtonian expansion,
	\refeq{eq:tidal_k2L}.
	
	
	\subsection{Remnant Properties}
	\label{sec:remnant_properties}
	
	Remnants in our simulations are characterised by a light accretion disc
	surrounding a spinning \ac{BH} formed $\lesssim 0.5~\rm ms$ after the merger. In
	the following we present the properties of both as extracted from our simulations.
	
	\subsubsection{Accretion disc}\label{sec:disc}

	\begin{figure}
		\includegraphics{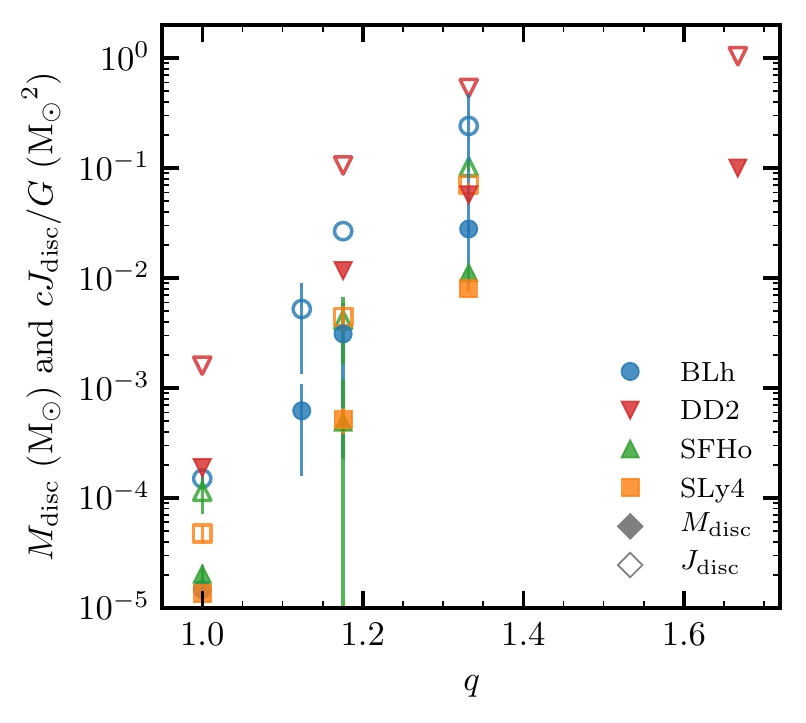}
		\caption{Disc mass (filled markers) and angular momentum (empty markers) at $4 -
			7$ ms after merger for SR simulations. Mass and angular momentum increase with
			the mass ratio. The trends suggest a link between mass and angular momentum
			since $c J_{\rm disc}/ G \sim (8 - 10) M_{\odot} \, M_{\rm disc}$. Errors are
			estimated as $|{\rm SR} - {\rm LR}|$ when the LR is available.}
		\label{fig:disc_mass_spin_q}
	\end{figure}
	
	During the last few orbits, the disc starts to form because of the tidal
	interaction between the two stars. In high-mass binaries resulting in prompt BH
	formation, the tidal interaction that occurs before and at merger is the major
	source of the disc. A few ms after merger the disc mass and angular momentum
	reach a quasi-steady phase, and slowly decrease until the end of the simulation.
	
	In \reffig{fig:disc_mass_spin_q}, we report the mass (filled markers) and
	angular momentum (unfilled markers) of the discs once they have reached their
	quasi-steady phase (i.e. $\sim 5 - 7$ ms after merger), computed as the integral
	of mass and angular momentum densities\footnote{This approach assumes that the metric is axisymmetric.} extracted from our simulations. The masses (angular
	momenta) span a broad range from $\sim 10^{-5} \Msun$ to $0.1 \Msun$ $(10^{-4}
	\; - \; 1 \; \Msun^2)$ depending on the BNS parameters. Both the disc mass and
	angular momentum increase as a function of the mass ratio $q$. We find that the
	increase is more pronounced for stiffer \acp{EOS}, where the tidal interaction
	is more efficient due to the larger $\tilde \Lambda$. For example, considering
	the trend for fixed $q=1.33$, the DD2 simulation ($\tilde \Lambda = 254$) leads
	to the formation of a disc twice more massive than the one formed in the BLh
	simulation ($\tilde \Lambda = 135$) and roughly six times more massive than
	those in the SFHo ($\tilde \Lambda = 105$) and SLy4 ($\tilde \Lambda = 93$)
	simulations. The errors on the disc mass, estimated when both resolutions are
	available as the absolute semi-difference between the SR and LR are in the range
	25-40 per cent for very light discs and get smaller ($\sim 3$ per cent) as the
	disc mass increases above $10^{-3}\Msun$. Resolution effects are higher for the
	BLh simulation with $q=1.18$, for which the disc mass of the LR simulation is
	$\sim 14$ times larger than the SR one. Despite efforts, we did not find the
	origin of such difference.
	
	\begin{figure}
		\includegraphics{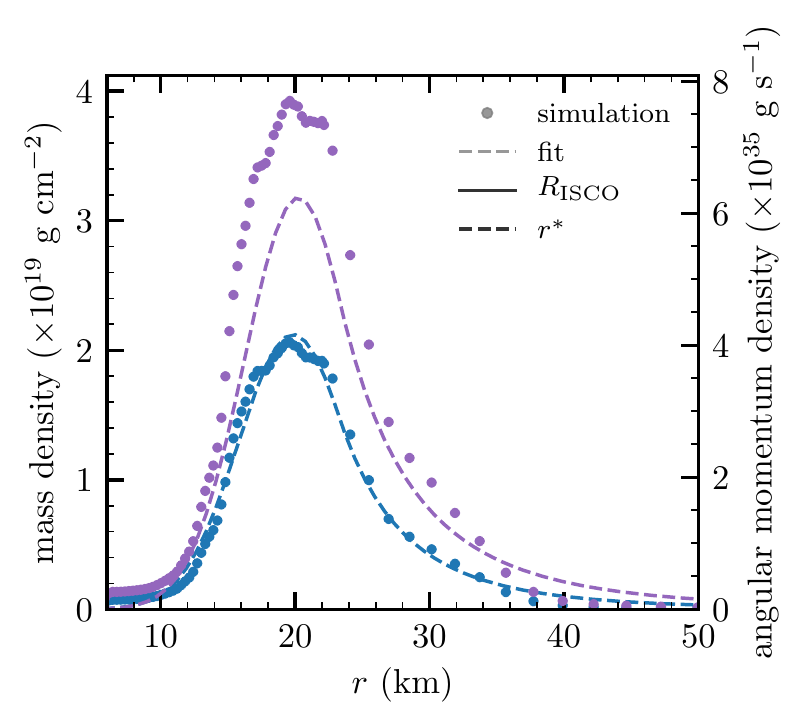}
		\caption{Disc's radial density (blue points, left $y$-axis) and radial
			angular momentum density (purple points, right $y$-axis) for the BNS with
			BLh \ac{EOS} and $q=1.33$. The blue dashed line is $\sigma(r)$ fitted on
			the numerical data, while the purple dashed line is the corresponding
			Keplerian angular momentum density. The vertical dashed line is the
			boundary between the Gaussian and the power-law $r^*$ in
			\refeq{eq:disc_radial_density}. The vertical solid line is $R_{\rm
				ISCO}$.}
		\label{fig:mass_ang_fit}
	\end{figure}
	\reffig{fig:disc_mass_spin_q} suggests a correlation between the mass and the
	angular momentum of the disc, i.e., $J_{\rm disc} \sim (8 - 10) M_{\odot} \, G
	M_{\rm disc} /c$, possibly independent from the \ac{EOS} and mass ratio. Stated
	differently, the mean specific angular momentum of the disc is (roughly)
	constant:
	$J_{\rm disc}/ M_{\rm disc} \sim (8 - 10) M_{\odot} \, G /c$.
	
	To provide a possible explanation, we consider the radial density distributions,
	$\sigma(r) = \int d\phi dz~\rho(r,\phi,z)$, as obtained from our numerical
	simulations, and we approximate it with a Gaussian peak smoothly connected to a
	radial power-law:
	\begin{equation}
	\label{eq:disc_radial_density}
	\sigma(r) =
	\begin{cases}
	b \exp\left( - \dfrac{(r-r_{\rm peak})^2}{2 s^2} \right) & 0 \leq r \leq r^* \\
	\sigma_0 \left( \dfrac{r}{r^*} \right)^{-\alpha} & r > r^*
	\end{cases}
	\end{equation}
	where $b$, $r_{\rm peak}$, $s$ and $\alpha$ are fitted against the actual radial
	density distribution in our simulations, while $\sigma_0$ and $r^*$ are fixed
	requiring $\sigma(r)$ to be differentiable in $r^*$. The parameter values and
	the quality of the fit are described in \refapp{app:keplerian}. Additionally, we
	assume a Keplerian angular velocity profile, $\omega_{\rm kep}(r) = \sqrt{G
		M_{\rm BH}/r^3}$, inside the disc. The mass and angular momentum of the
	resulting Keplerian disc are:
	\begin{align} \label{eq:mass_J_keplerian}
	&M_{\rm disc}^{\rm kep} = \int_{0}^{\infty} r \sigma(r) {\rm d}r,
	&&J_{\rm disc}^{\rm kep} = \int_{0}^{\infty} r^3 \sigma(r) \, \omega_{\rm kep}(r) {\rm d}r.
	\end{align}
	In \reffig{fig:mass_ang_fit}, we show the result of the fit for $\sigma(r)$
	(blue dashed line) on the numerical one (blue dots) for the simulation with the
	BLh \ac{EOS} and $q=1.33$. We also show the radial angular momentum density from
	the numerical simulation (purple points) and the corresponding Keplerian
	analogue computed from \refeq{eq:mass_J_keplerian} with the fitted $\sigma(r)$
	(purple dashed line). We found that $J_{\rm disc}^{\rm kep} \lesssim J_{\rm
		disc}$, usually within $30$ per cent over more than two orders of magnitudes
	in $J_{\rm disc}$.
	We excluded the discs of equal mass \ac{BNS} from this analysis since
	they are very light and $40-100$ per cent of their mass is inside the \ac{ISCO} predicted according to the \ac{BH} properties. Such discs will be accreted by the \ac{BH} on the viscous timescale.
	Given Eqs.~(\ref{eq:disc_radial_density})-(\ref{eq:mass_J_keplerian}), the ratio
	between $J_{\rm disc}^{\rm kep}$ and $M_{\rm disc}^{\rm kep}$ can be written as
	(see \refapp{app:keplerian} for a derivation):
	\begin{equation}\label{eq:keplerian_ratio}
	\frac{J^{\rm kep}_{\rm disc}}{M^{\rm kep}_{\rm disc}} = \left( \eta \, \frac{\alpha - 2}{\alpha - 5/2} \sqrt{\frac{M_{\rm BH}}{\msun} \frac{2 r^*}{R^{\rm Sch}_{\odot}}} \right)~\frac{G \Msun}{c} \, ,
	\end{equation}
	where $\eta$ is defined as in \refeq{eq:eta} and varies between 0.78 and 0.90
	with average $0.83$ in our numerical simulations, $R_{\odot}^{\rm Sch}$ is the
	Schwarzschild radius of the Sun, $r^*$ is such that $ 21{\rm km} \lesssim r^*
	\lesssim 40{\rm km}$, while $M_{\rm BH} \approx 3.21-3.26 \msun$ (see
	\refsec{sec:BH}). The parameter which is subject to more significant variation
	is $\alpha \in [4.0, 13.9]$ whose average is $7.5$ (see \refapp{app:keplerian}
	for the values of $\alpha$ and $r^*$). Inserting these ranges of values in
	\refeq{eq:keplerian_ratio}, one obtains $J^{\rm kep}_{\rm disc}/M^{\rm kep}_{\rm
		disc} \sim 6-9~\Msun$ with average of $7.3~\Msun$, in agreement within
	$\approx 83$ per cent with the average $\langle J_{\rm disc} / M_{\rm disc}
	\rangle = 8.8~\Msun$ obtained by our numerical simulations.

	\subsubsection{Black hole}\label{sec:BH}
	
	\begin{figure}
		\includegraphics{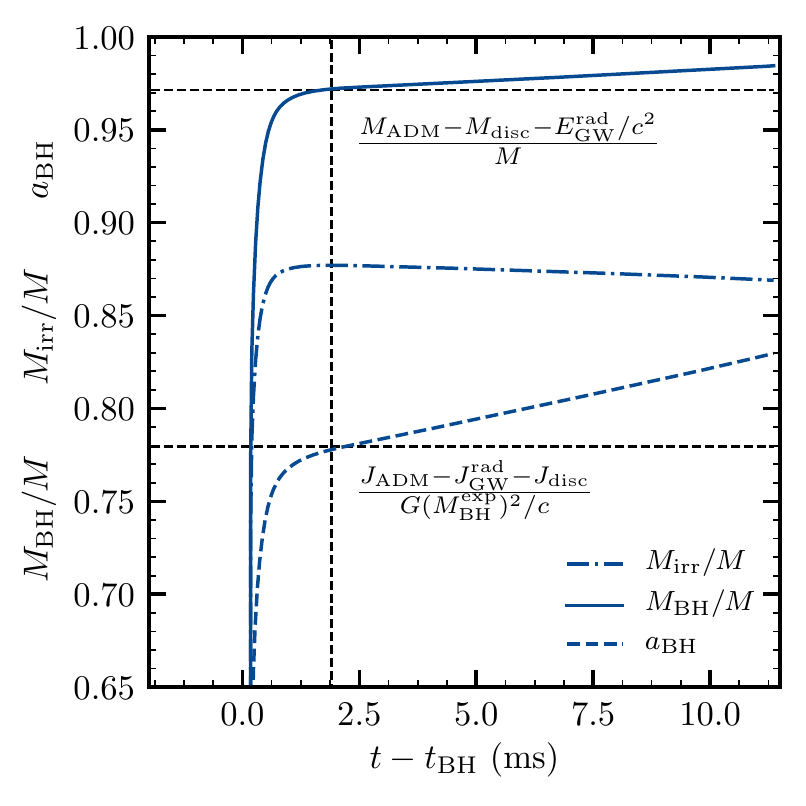}
		\caption{Evolution of the normalised BH irreducible mass $M_{\rm irr}/M$,
			gravitational mass $M_{\rm BH}/M$ and dimensionless spin parameter $a_{\rm
				BH}$ for a SR simulation based on the BLh \ac{EOS} with $q=1.33$. Horizontal
			dashed lines represent the expected values for the gravitational mass $(M_{\rm
				ADM} - E_{\rm GW}^{\rm rad} - M_{\rm disc})/M$ and the spin parameter
			$(J_{\rm ADM} - J_{\rm GW}^{\rm rad} - J_{\rm disc} ) / (M_{\rm BH}^{\rm
				exp})^2$. Vertical dashed lines indicate the time at which the irreducible
			mass starts to decrease and the corresponding value on the plotted line.}
		\label{fig:BLh_q133_irr_grav_spin}
	\end{figure}
	
	In \reffig{fig:BLh_q133_irr_grav_spin} we report the BH irreducible and
	gravitational masses,
	and the dimensionless spin parameter
	as a function of time after the BH formation for the BLh simulation at SR with
	$q=1.33$. We see that all the three quantities increase abruptly as the AH
	finder detects the apparent horizon. The horizontal dashed lines indicate the
	expected values $M^{\rm exp}_{\rm BH}$ and $a^{\rm exp}_{\rm BH}$, while the
	vertical dashed line indicates the time at which the irreducible mass reaches
	its maximum value (a few ms after the BH formation). Although $M_{\rm irr}$ is
	expected to remain constant or to increase, we find that after having reached
	the maximum it starts to slowly decrease. We attribute this behaviour to
	numerical and discretisation errors in tracing the AH location. While the AH
	shrinks, $M_{\rm BH}$ and $a_{\rm BH}$ continue to increase without reaching
	saturation. Matter accretion from the disc is not sufficient to explain this
	growth. The rise of $M_{\rm BH}$ after the maximum of $M_{\rm irr}$ is due to
	the continuous increase of the BH spin, which is an artefact of our simulations.
	Due to these uncertainties, we decide to focus on the gravitational mass and
	spin parameter of the BH at the moment when the irreducible mass is maximum.
	
	In \reftab{tab:rem_props} we report the gravitational mass $M_{\rm BH}$ and the
	spin parameter $a_{\rm BH}$ of the BH computed on the basis of the latter
	definition. To give more conservative values of the BH properties, we report
	also the time averages of the BH mass, $\langle M_{\rm BH} \rangle$, and spin
	parameter, $\langle a_{\rm BH} \rangle$, over the first $7~\rm ms$ after the
	time at which $M_{\rm irr}$ is maximum. We report the available data obtained by
	SR simulations and we estimate the uncertainties (when available) as the
	semi-difference with respect to the data from the corresponding LR simulations
	when available. In the case of simulations employing the BLh or SFHo \ac{EOS},
	the AH is resolved by the AH finder and the BH properties can be analysed with
	appropriate accuracy. More quantitatively, $M_{\rm BH}$ and $a_{\rm BH}$ differ
	from the respective expected values less than 1 per cent. On the other hand, the
	AH finder was unable to detect the AH for the simulations employing the DD2 or
	SLy4 \ac{EOS}. In these cases we decided not to report the corresponding values
	in \reftab{tab:rem_props}.
	
	\begin{figure}
		\includegraphics{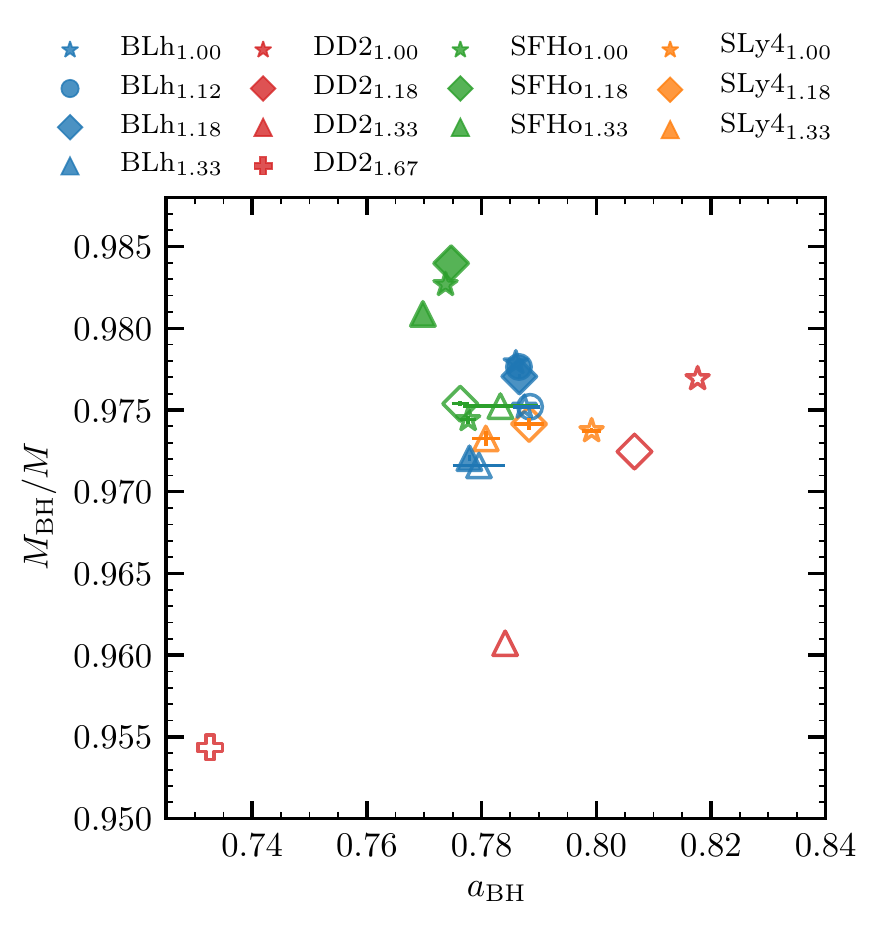}
		\caption{$M_{\rm BH}/M$ and dimensionless spin parameter $a_{{\rm BH}}$
			distribution for the SR simulations of this work. Filled markers represent
			the values computed by the AH finder, while empty markers represent the
			expected ones. Errors are computed as the absolute semi-difference between
			SR and LR when available. For the filled markers errors are smaller than
			the symbol size.}
		\label{fig:bh_trend}
	\end{figure}
	
	Regarding the dependence of the BH properties on the initial binary parameters,
	the final outcome depends mostly on two effects. On one hand, energy and
	angular momentum are extracted from the central object via the ejection of
	matter and the formation of a remnant disc. On the other hand, GWs carry energy
	and angular momentum away. Both these effects reduce at the same time $M_{\rm
		BH}$ and $J_{\rm BH}$. Since $J_{\rm disc} \approx 10~\msun~G/c~M_{\rm disc}$,
	the formation of a massive disc is particularly efficient in reducing the BH
	angular momentum, and ultimately also the spin parameter since the variation of
	$a_{\rm BH}^{\rm exp}$ due to the disc formation only becomes $\left. \delta
	a_{\rm BH}^{\rm exp} \right|_{\rm disc} \approx (2a_{\rm BH}^{\rm exp}-10
	\msun/M_{\rm BH}^{\rm exp} ) \delta M_{\rm disc}/M_{\rm BH}^{\rm exp} \sim -
	0.468 ~ \delta M_{\rm disc}/(\msun) $. As visible in \reffig{fig:bh_trend},
	(quasi) equal mass binary simulations employing the DD2 \ac{EOS} have the
	largest spin parameters, since their symmetric character produces a smaller disc
	mass, while their larger $\kappa_2^{\rm L}$ implies a lower GW emission.
	However, very asymmetric binaries employing the same \ac{EOS} produce massive
	discs reducing efficiently both $M_{\rm BH}$ and $a_{\rm BH}$. A similar, but
	less significant effect, is also observed for simulations employing the BLh and
	SFHo \acp{EOS}. For simulations employing the SLy \ac{EOS} (whose discs are
	usually the lightest), $a_{\rm BH}$ decreases with $q$, while $M_{\rm BH}/M$
	stays roughly constant. Focusing on the (quasi-)equal mass simulations using the
	BLh, SFHo or SLy4 \ac{EOS}, the removal of mass and angular momentum through the
	disc formation becomes subdominant, while the dominant process is the GW
	emission. More symmetric binaries modelled with the SLy4 \ac{EOS} (corresponding
	to lower values of $\kappa_2^{\rm L}$), have indeed the smallest BH masses.
	
	\subsection{Dynamical Ejecta}
	\label{sec:dynamical_ejecta}
	\begin{table*}
		\caption{Dynamical ejecta properties for each simulation. $\md$ is the total
			mass of the ejecta; $\sdtd$ and $\sdpd$ are the mass-weighted standard
			deviation of the polar and azimuthal angle, respectively; $\mvd$, $\myd$ and $\msd$ are the median values of the electron fraction, speed and entropy distributions. The last column is the ratio $X_s\equiv M_{\rm ej}^{\rm shocked} / \md$, where the shocked and tidal ejecta are defined as the components with entropy respectively above and below the threshold of $10\ k_{\rm B}~{\rm baryon}^{-1}$. The subscript and superscript numbers indicate the 15 and 75 percentile around the median of the
			respective quantity.}
		\label{tab:ejecta}
		\begin{tabular}{cc|cccccccc}
			\hline\hline
			EOS & $q$ & 
			\makecell{
				Resolution \\ $\,$
			} &
			\makecell{
				$\md$ \\ $[10^{-4} \Mo]$
			} &
			\makecell{
				$\sdtd$ \\ $\,$
			} &
			\makecell{
				$\sdpd$ \\ $\,$
			} &
			\makecell{
				$\mvd$ \\ $[c]$
			} &
			\makecell{
				$\myd$ \\ $\,$
			} &
			\makecell{
				$\msd$ \\ $[k_{\rm B}~{\rm baryon}^{-1} ]$
			} &
			\makecell{
				$X_s$ \\ $\,$
			}
			\\
			\hline
			BLh & 1.0 & 
			\makecell{    
				SR \\ LR
			} &
			\makecell{    
				0.002 \\ 0.023
			} &
			\makecell{    
				- \\ -
			} &
			\makecell{    
				- \\ -
			} &
			\makecell{    
				- \\ -
			} &
			\makecell{    
				- \\ -
			} &
			\makecell{    
				- \\ -
			} &
			\makecell{    
				- \\ -
			} \\   
			BLh & 1.12 & 
			\makecell{    
				SR \\ LR 
			} &
			\makecell{    
				0.039 \\ 0.090
			} &
			\makecell{    
				- \\ -
			} &
			\makecell{    
				- \\ -
			} &
			\makecell{    
				- \\ -
			} &
			\makecell{    
				- \\ -
			} &
			\makecell{    
				- \\ -
			} &
			\makecell{    
				- \\ -
			} \\
			BLh & 1.18 & 
			\makecell{    
				SR \\ LR
			} &
			\makecell{    
				0.164 \\ 0.182
			} &
			\makecell{    
				21.3 \\ 23.3
			} &
			\makecell{    
				82.0 \\ 89.8
			} &
			\makecell{    
				$0.24 \err{0.12}{0.08}$ \\ $0.21 \err{0.10}{0.07}$
			} &
			\makecell{    
				$0.21 \err{0.08}{0.07}$ \\ $0.25 \err{0.07}{0.04}$
			} &
			\makecell{    
				$18.1 \err{11.6}{39.4}$ \\ $41.2 \err{31.5}{55.4}$
			} &
			\makecell{    
				0.78 \\ 0.94
			} \\
			BLh & 1.33 & 
			\makecell{    
				SR \\ LR
			} &
			\makecell{    
				0.508 \\ 0.959
			} &
			\makecell{    
				18.2 \\ 20.7
			} &
			\makecell{    
				74.0 \\ 78.6
			} &
			\makecell{    
				$0.27 \err{0.14}{0.10}$ \\ $0.29 \err{0.15}{0.10}$
			} &
			\makecell{    
				$0.17 \err{0.5}{0.9}$ \\ $0.16 \err{0.5}{0.14}$
			} &
			\makecell{    
				$9.71 \err{4.21}{17.4}$ \\ $12.3 \err{6.87}{22.0}$
			} &
			\makecell{    
				0.61 \\ 0.63
			}
			\\
			\hline
			DD2 & 1.0 & 
			\makecell{    
				SR \\ LR
			} &
			\makecell{    
				0.586 \\ 0.416
			} &
			\makecell{    
				26.3 \\ 23.8
			} &
			\makecell{    
				95.1 \\ 92.1
			} &
			\makecell{    
				$0.28 \err{0.12}{0.09}$ \\ $0.32 \err{0.08}{0.06}$
			} &
			\makecell{    
				$0.27 \err{0.06}{0.04}$ \\ $0.29 \err{0.05}{0.03}$
			} &
			\makecell{    
				$33.2 \err{18.3}{38.8}$ \\ $47.1 \err{31.4}{42.4}$
			} &
			\makecell{    
				1.00 \\ 1.00 
			} \\
			DD2 & 1.18 & 
			\makecell{    
				SR \\ LR
			} &
			\makecell{    
				7.16 \\ 9.67
			} &
			\makecell{    
				21.4 \\ 18.1
			} &
			\makecell{    
				122 \\ 87.3
			} &
			\makecell{    
				$0.27 \err{0.14}{0.10}$ \\ $0.27 \err{0.15}{0.11}$
			} &
			\makecell{    
				$0.17 \err{0.06}{0.05}$ \\ $0.19 \err{0.08}{0.06}$
			} &
			\makecell{    
				$10.28 \err{4.12}{7.18}$ \\ $9.36 \err{3.80}{5.42}$
			} &
			\makecell{    
				0.57 \\ 0.63
			} \\
			DD2 & 1.33 & 
			\makecell{    
				SR \\ LR
			} &
			\makecell{    
				4.00 \\ 3.94
			} &
			\makecell{    
				17.3 \\ 21.7
			} &
			\makecell{    
				76.6 \\ 80.7
			} &
			\makecell{    
				$0.23 \err{0.11}{0.08}$ \\ $0.19 \err{0.11}{0.10}$
			} &
			\makecell{    
				$0.15 \err{0.05}{0.05}$ \\ $0.13 \err{0.05}{0.8}$
			} &
			\makecell{    
				$9.38 \err{3.66}{3.64}$ \\ $9.34 \err{3.29}{5.15}$
			} &
			\makecell{    
				0.65 \\ 0.52
			} \\
			DD2 & 1.67 & 
			\makecell{    
				SR \\ LR
			} &
			\makecell{    
				4.05 \\ 6.20
			} &
			\makecell{    
				11.1 \\ 13.0
			} &
			\makecell{    
				103 \\ 95.8
			} &
			\makecell{    
				$0.20 \err{0.14}{0.14}$ \\ $0.13 \err{0.8}{0.13}$
			} &
			\makecell{    
				$0.10 \err{0.07}{0.03}$ \\ $0.06 \err{0.03}{0.08}$
			} &
			\makecell{    
				$5.66 \err{1.87}{4.27}$ \\ $6.15 \err{3.33}{3.70}$
			} &
			\makecell{    
				0.29 \\ 0.37
			}
			\\
			\hline
			SFHo & 1.0 & 
			\makecell{    
				SR \\ LR
			} &
			\makecell{    
				0.023 \\ 0.033
			} &
			\makecell{    
				- \\ -
			} &
			\makecell{    
				- \\ -
			} &
			\makecell{    
				- \\ -
			} &
			\makecell{    
				- \\ -
			} &
			\makecell{    
				- \\ -
			} &
			\makecell{    
				- \\ -
			} \\
			SFHo & 1.18 & 
			\makecell{    
				SR \\ LR
			} &
			\makecell{    
				0.071 \\ 0.151
			} &
			\makecell{    
				- \\ 24.5
			} &
			\makecell{    
				- \\ 90.6
			} &
			\makecell{    
				- \\ $0.22 \err{0.10}{0.07}$
			} &
			\makecell{    
				- \\ $0.26 \err{0.04}{0.03}$
			} &
			\makecell{    
				- \\ $72.3 \err{53.1}{51.3}$
			} &
			\makecell{    
				- \\ 0.97
			} \\
			SFHo & 1.33 & 
			\makecell{    
				SR \\ LR
			} &
			\makecell{    
				0.603 \\ 1.87
			} &
			\makecell{    
				12.7 \\ 13.1
			} &
			\makecell{    
				68.8 \\ 85.0
			} &
			\makecell{    
				$0.26 \err{0.13}{0.10}$ \\ $0.32 \err{0.16}{0.10}$ 
			} &
			\makecell{    
				$0.13 \err{0.06}{0.04}$ \\ $0.13 \err{0.05}{0.05}$
			} &
			\makecell{    
				$7.55 \err{3.30}{4.97}$ \\ $6.45 \err{2.50}{5.08}$
			} &
			\makecell{    
				0.37 \\ 0.32
			}
			\\
			\hline
			SLy4 & 1.0 & 
			\makecell{    
				SR \\ LR
			} &
			\makecell{    
				0.030 \\ 0.024
			} &
			\makecell{    
				- \\ -
			} &
			\makecell{    
				- \\ -
			} &
			\makecell{    
				- \\ -
			} &
			\makecell{    
				- \\ -
			} &
			\makecell{    
				- \\ -
			} &
			\makecell{    
				- \\ -
			} \\
			SLy4 & 1.18 & 
			\makecell{    
				SR \\ LR
			} &
			\makecell{    
				0.055 \\ 0.114
			} &
			\makecell{    
				- \\ 21.4
			} &
			\makecell{    
				- \\ 79.5
			} &
			\makecell{    
				- \\ $0.22 \err{0.10}{0.10}$
			} &
			\makecell{    
				- \\ $0.24 \err{0.06}{0.05}$
			} &
			\makecell{    
				- \\ $38.1 \err{31.4}{97.5}$
			} &
			\makecell{    
				- \\ 0.79
			} \\
			SLy4 & 1.33 & 
			\makecell{    
				SR \\ LR
			} &
			\makecell{    
				2.29 \\ 1.12
			} &
			\makecell{    
				9.0  \\ 14.6
			} &
			\makecell{    
				71.5 \\ 70.8
			} &
			\makecell{    
				$0.40 \err{0.20}{0.12}$ \\ $0.30 \err{0.14}{0.10}$
			} &
			\makecell{    
				$0.10 \err{0.02}{0.03}$ \\ $0.12 \err{0.5}{0.09}$
			} &
			\makecell{    
				$5.48 \err{3.15}{1.82}$ \\ $7.40 \err{4.44}{8.42}$
			} &
			\makecell{    
				0.22 \\ 0.49
			}
			\\
			\hline\hline
		\end{tabular}
	\end{table*}

\begin{figure*}
		\includegraphics{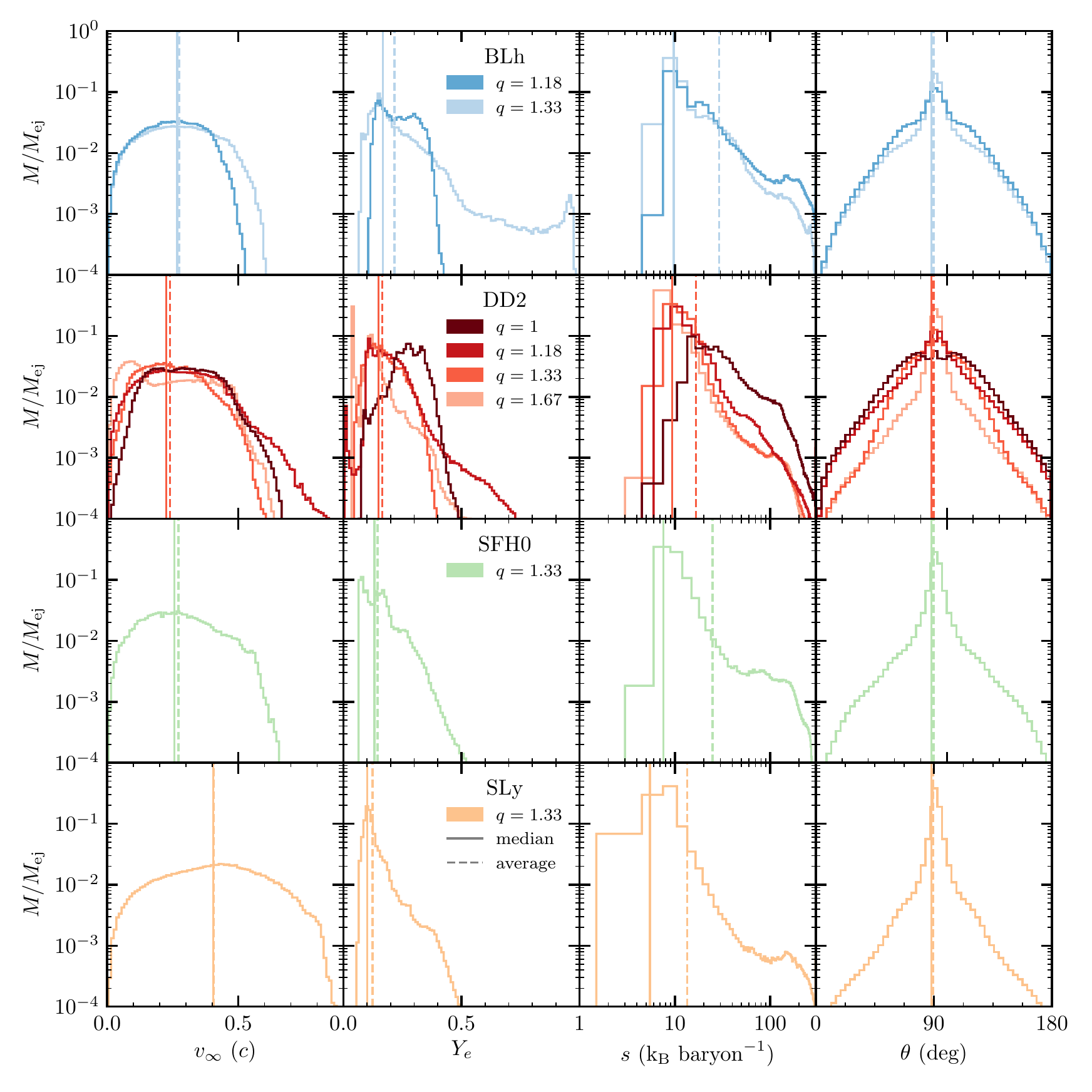}
		\caption{Histograms of the dynamical ejecta. From the first to the last column:
			velocity at infinity $\vd$, electron fraction $\yd$, entropy per baryon $s$
			and polar angle $\td$. Each row represents a different \ac{EOS}. From the
			first to the last line: BLh, DD2, SFHo, SLy4. As a representative case, we
			represent the median and the average values of all quantities for the $q =
			1.33$ cases as vertical solid and dashed lines, respectively. The high $Y_e$ tail in the BLh, $q=1.33$ case is not robust due to the finite size of the \ac{EOS} tables not extending above $Y_e = 0.6$.}
		\label{fig:ejecta_histograms}
	\end{figure*}

	In \reftab{tab:ejecta}, we present the properties of the dynamical ejecta as
	extracted from our simulations, namely the mass of the ejecta, $\md$; the
	\ac{SD} of the polar ($\theta \in [0^\circ,180^\circ]$) and azimuthal ($\phi \in
	[0^\circ, 360^\circ]$, see \refapp{sec:appendix_rms} for more details on its
	calculation) angular distributions, $\sdtd$ and $\sdpd$, respectively; the
	median of the distribution of the velocity at infinity, $\mvd$, of the electron
	fraction, $\myd$, and of the entropy per baryon, $\msd$. The last column refers
	to the fraction of shocked ejecta $X_s$, defined as the fraction of the ejecta
	whose entropy is larger than $10 \, k_{\rm B}~{\rm baryon^{-1}}$. We report the
	values for both SR and LR simulations accompanied by the 15-75
	percentile range around the median computed from the respective mass-weighted histogram. We do not report the ejecta properties when
	$\md < 10^{-5} \Msun$, since the properties of such a small amount of ejected
	matter cannot be trusted due to numerical uncertainties.
	Additionally, in \reffig{fig:ejecta_histograms}, we present mass histograms of
	the $\vd$, $\yd$, $\sd$ and $\td$ distributions for simulations at SR for which
	$M_{\rm ej}\geq 10^{-5} \Msun$. The vertical solid (dashed) lines represent the
	medians (average) of the ejecta properties for the $q = 1.33$ cases, taken as
	representative case. While the difference between mean and median is small or
	even negligible for the velocity and the electron fraction, a significant
	difference is clear in the entropy distribution.
	
	\begin{figure}
		\includegraphics{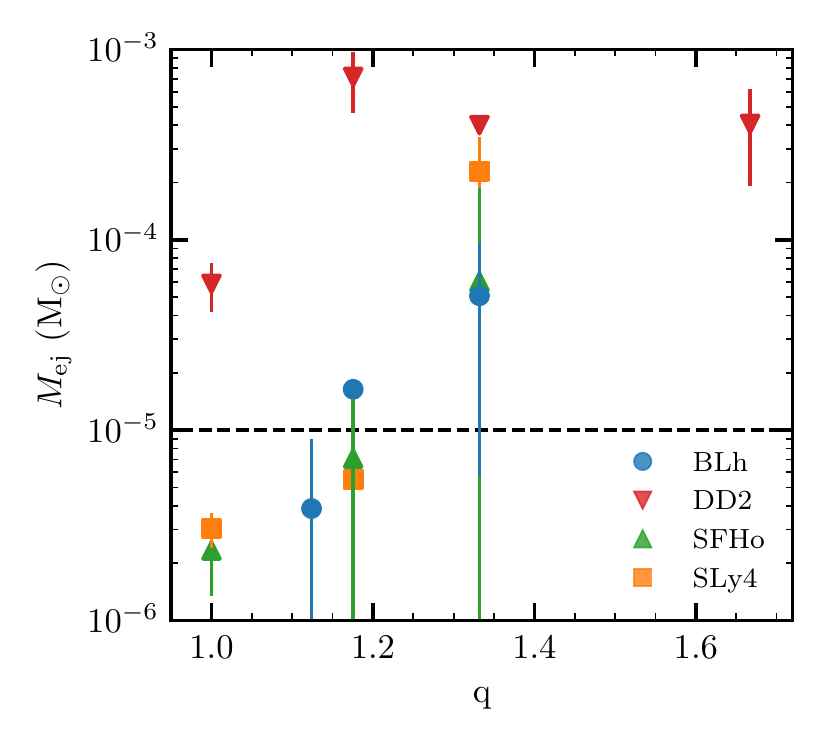}
		\caption{Dynamical ejecta mass as a function of the mass ratio $q$ of the
			binary. Different symbols denote numerical simulations with different
			\ac{EOS}. Simulations with $M_{\rm ej}<10^{-6} \; \Msun$ have been
			excluded, while only ejecta with $M_{\rm ej}>10^{-5} \; \Msun$ is trusted
			due to numerical uncertainties. Errors are computed as the absolute difference between SR and LR values.}
		\label{fig:dudi_ej_comparison}
	\end{figure}
	
	The ejecta mass ranges from values smaller than $10^{-5}\Msun$ up to $\sim 6
	\times 10^{-4} \Msun$, increasing with the mass ratio $q$ and the stiffness of
	the \ac{EOS}, as visible in \reffig{fig:dudi_ej_comparison}. For asymmetric
	systems ($q \neq 1$) and stiffer \acp{EOS}, the tidal interaction is more
	efficient in deforming the secondary NS and the resulting merger dynamics is
	more effective in expelling matter from its tidal tails \citep[see
	e.g.][]{Hotokezaka:2012ze, Bauswein:2013yna, Sekiguchi:2015dma,
		Rosswog:2015nja, Lehner:2016lxy, Dietrich:2016hky, Bernuzzi:2020txg}.
	Simulations employing the DD2 \ac{EOS} exhibit a deviation from this trend at higher mass ratios ($q = 1.33, \, 1.67$), for which the value of the ejecta mass saturates or even tends to decrease, similarly to what found in \citet{Dudi:2021abi} (see \refsec{sec:discussion}). We speculate that the ejection process at high $q$'s is more sensitive to usually subdominant effects, including the detailed behaviour of the NS radius and of $\tilde{\Lambda}$, see \reffig{fig:TOV} and \reftab{tab:sim}. For the latter quantity, for high-$q$ BNSs, models employing the DD2 show a decreasing $\tilde \Lambda$ (see \reftab{tab:sim}). It suggest that for asymmetric enough BNS ($q \gtrsim 1.2$ in our case), if an additional increase of the asymmetry is not accompanied by and increase of $\tilde \Lambda$, the ejecta mass can saturate or even decrease. More simulations at higher resolutions are needed to confirm the robustness of this trend.
	
	The SD of the geometrical angles gives an indication of the spatial distribution
	of the ejected matter. We find that the ejecta spread over the whole space, but
	it is mostly concentrated close to the equator, with an opening angle $2 \sdtd$
	that varies across the range $18^{\circ} - 54^{\circ}$, depending on the binary
	properties and where higher values correspond to more symmetric binaries. This
	can be understood since the tidal interaction tends to distribute matter along
	the orbital plane. The SD of the azimuthal angle $\sdpd$ is related to the
	rotational symmetry of the dynamical ejecta around the orbital axis. For a mass
	distribution uniform in $\phi$ and centred in $180^{\circ}$ with symmetric
	support on $2 \alpha \in [0, 360^{\circ}]$, we expect a SD of $\sdpd =
	(\sqrt{3}/3)~\alpha \approx 52^{\circ} (\alpha/90^{\circ})$. The values of
	$\sdpd$ obtained in our simulations range within $65^{\circ} - 96^{\circ}$ and
	are compatible with a uniform distribution centred in $180^{\circ}$ with support
	on $\sim 225^{\circ} - 360^{\circ}$ respectively, where higher values correspond
	to equal-mass systems. This indicates that the dynamical ejecta expelled by
	symmetric binaries is distributed over the whole azimuthal angle, while the
	anisotropy increases with $q$ \citep[see
	e.g.][]{Bovard:2017mvn,Radice:2018pdn,Bernuzzi:2020txg}.
	
	The distribution of the radial velocity at infinity has $\mvd$ ranging from
	$\sim 0.2 \, c$ to $\sim 0.4 \, c$, with fast tails reaching $\sim 0.6-0.9 \, c$
	for the highest mass ratios.
	The median of the electron fraction distribution is always smaller than $0.3$
	and is lower for higher mass ratios: tidal interaction ejects cold neutron rich
	material only marginally subject to composition reprocessing from positron and
	neutrino captures
	\citep[e.g.][]{Wanajo:2014wha,Sekiguchi:2015dma,Perego:2017wtu,Martin:2017dhc}.
	Finally, the entropy per baryon has a distribution with a marked peak at
	relatively low entropy, between $\sim 5 \; k_{\rm B}~{\rm baryon}^{-1}$ and
	$\sim 20 \; k_{\rm B}~{\rm baryon}^{-1}$, and a slow decrease towards higher
	entropy, with medians that in the SR cases range between $\sim 5 \; k_{\rm
		B}~{\rm baryon}^{-1}$ and $\sim 18 \; k_{\rm B}~{\rm baryon}^{-1}$ (with the
	only exception of the $q=1$ simulation employing the DD2 \ac{EOS}, and more
	often $ \lesssim 10~k_{\rm B}~{\rm baryon}^{-1}$). All the entropy distributions
	show a second peak around $\sd \sim 120 \; k_{\rm B}~{\rm baryon}^{-1}$
	whose relative importance decreasing with $q$ and with the stiffness of the
	\ac{EOS}, ranging approximately between $10^{-2}$ and $10^{-3}$. This
	high-entropy component reflects the presence of a shocked fraction of the ejecta
	coming from the collisional interface of the two NSs (see
	Sec.~\ref{sec:merger_dynamics} and Fig.~\ref{fig:dens_entr}). We expect this
	component to be present also in BNS mergers characterised by lower total masses
	(and often not resulting in a prompt collapse), in which the total amount of
	ejected matter is typically larger than what found in our simulations.
	The compositional properties of the dynamical ejecta show distributions
	comparable to what studied in \citet{Most:2020exl} for the case of an
	irrotational binary, with similar fast-tail, high ye and high entropy
	components.
	\begin{figure*}
		\includegraphics{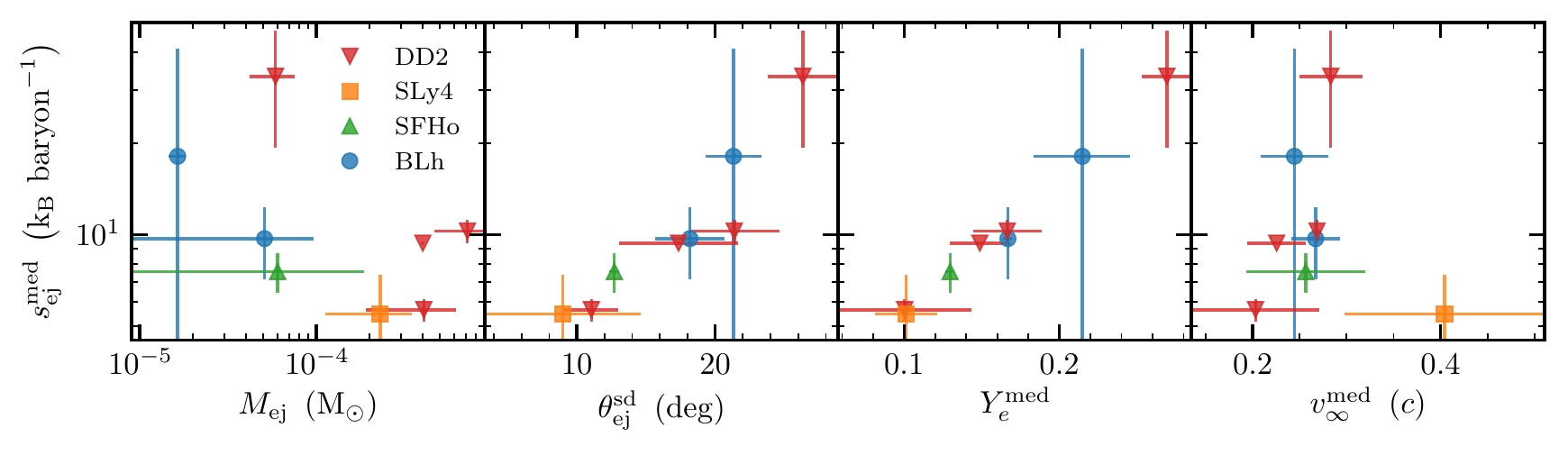}
		\caption{Correlation of the ejecta mass $M_{ej}$, standard deviation of the
			polar angle $\sdtd$, median of the electron fraction $\myd$ and median of
			the velocity at infinity $\mvd$ with the median of the entropy $\msd$.
			Uncertainties are estimated as the absolute difference between SR and LR simulations, while SR values are used to represent the points. The
			simulations with higher mass ratios have higher values of the ejected
			mass.}
		\label{fig:ejecta_corr_s}
	\end{figure*}
	
	In the analysis outlined above, we have found that many properties of the
	ejected matter correlate with $q$ and with the \ac{EOS} stiffness. We now
	explicitly explore correlations among the different ejecta properties. In
	\reffig{fig:ejecta_corr_s}, we show $\md$, $\myd$ and $\sdtd$ as a function of
	$\msd$ for each BNS simulation producing more than $10^{-5}\Msun$ of dynamical
	ejecta. We recall that lower $\msd$ correspond to higher values of $q$. In the
	left panel we observe that $\md$ is larger for lower values of $\msd$ and it is
	usually greater for stiffer \acp{EOS}. In the two middle panels, we observe that
	both $\sdtd$ and $\myd$ increase almost linearly with the logarithm of the
	median of the entropy distribution. This confirms that the tidal interaction
	tends to distribute cold, low-entropy ejecta along the orbital plane. Only for
	simulations in which the shock-heated component is relevant (i.e., symmetric or
	nearly symmetric BNSs), the angular distribution of the ejecta departs
	significantly from the orbital plane, indicating that shocked matter spreads
	more over the solid angle. Similar results were found also for unequal-mass
	binaries that do not collapse promptly into a black hole. \citep[see
	e.g.][]{Bauswein:2013yna, Lehner:2016lxy, Dietrich:2016hky, Radice:2018pdn,
		Bernuzzi:2020txg, Nedora:2021eoj}. In the right panel, we study the
	correlations between the median of the entropy and the median of the velocity at
	infinity. In our simulations $\mvd$ decrease with $\msd$, indicating that higher
	mass ratios result in faster ejecta, contrary to what usually found in relation
	to systems characterised by smaller total masses. This could be indeed a
	peculiar property of very massive BNSs.
	
	\section{Nucleosynthesis and kilonova}
	\label{sec:nucleosynthesis and kilonova}
	
	\subsection{Nucleosynthesis}
	\label{sec:nucleosynthesis}
	\begin{figure*}
		\includegraphics{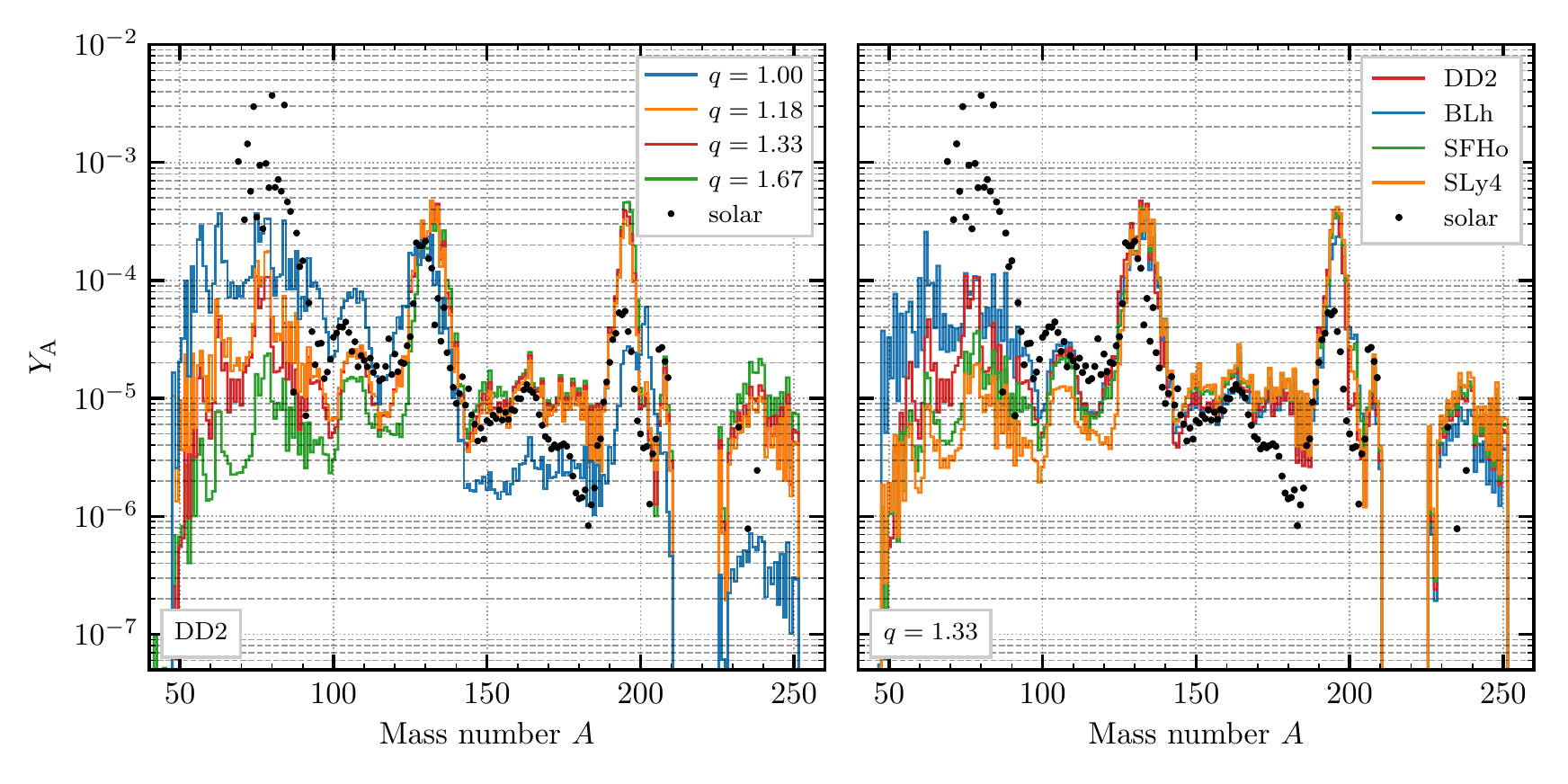}
		\caption{Nucleosynthesis pattern at $t=30$ years after the merger as a function
			of the mass number $A$. Left: comparison between relative abundances from
			simulations employing the DD2 \ac{EOS}. Right: comparison between relative
			abundances from \ac{NR} simulations with mass ratio $q=1.33$. Black dots represent
			the Solar $r$-process abundances, taken from \citet{Prantzos2020}. To guide
			the comparison, the Solar residuals are scaled in order to reproduce at
			$A=130$ the abundance of the simulation with $q=1.33$ and the DD2 \ac{EOS}.}
		\label{fig:nucleosynthesis_hist}
	\end{figure*}
	
	Using the procedure outlined in \refsec{subsec:nucleosynthesis calculations}, we
	compute nucleosynthesis yields for the dynamical ejecta of all our GW190425
	targeted simulations.
	In \reffig{fig:nucleosynthesis_hist}, we present nucleosynthesis yields for a
	subset of representative simulations at $t=30$ years after merger, superimposed
	to the Solar residual $r$-process abundances taken from \citet{Prantzos2020} as a useful point of reference. To
	guide the comparison between the different models, the Solar residuals are
	scaled in order to reproduce the abundance of the simulation with $q=1.33$ and
	the DD2 \ac{EOS} at $A=130$.
	
	Unequal-mass merger simulations employing the DD2 \ac{EOS} (left panel) robustly
	produce elements between the second and the third $r$-process peak, without
	showing any substantial difference between the various mass ratios. Relative
	abundances are comparable to the Solar residuals with a significant excess in
	the third peak height with respect to the height of the second peak, and a
	significant production of translead nuclei. On the other hand, $A\lesssim120$
	nuclei are systematically underproduced. A weak dependence on the value of the
	mass ratio is visible, with more asymmetric mergers producing on average a
	larger amount of heavy nuclei. These behaviours are expected given the prompt
	collapse of the central remnant into a BH, the tidal character of the ejection
	mechanism and the consequent absence of a significant high-$Y_e$ tail in the
	dynamical ejecta above a critical value $Y_e\gtrsim0.22$
	\citep[e.g.][]{Lippuner:2015gwa,Radice:2016dwd}, that is associated with the
	production of less than 10 per cent of the mass fraction of heavy nuclei above
	the second peak through an incomplete $r$-process.
	
	The situation changes significantly when considering the DD2 equal-mass case
	(blue line). In fact, the relative abundances of heavy $r$-process nuclei ($A
	\gtrsim 130$ and even more for $A\gtrsim140$) are less significant with respect
	to the unequal mass cases, while around the first peak the $q=1$ pattern is the
	largest and the closest one to the Solar abundances. This is consistent with the
	fact that, despite having a small total mass, the bulk of the ejecta $Y_e$
	distribution for the equal-mass case lies within the interval $0.20-0.40$ (see
	\reffig{fig:ejecta_histograms}).
	
	The right panel of \reffig{fig:nucleosynthesis_hist} shows, instead, the
	comparison between simulations characterised by the same mass ratio, namely
	$q=1.33$, but different \acp{EOS}. Since the mass ratio differs significantly
	from 1, the nucleosynthesis outcome is in all cases similar to what described
	for unequal-mass merger simulations in the comparison between the DD2
	simulations.
	All the curves are quite close to each other except around the first peak, where
	the spread between the various distributions becomes more evident and sensitive
	to the nuclear \ac{EOS}, with the largest (smallest) relative values for the
	abundances obtained for the BLh (SLy4) \ac{EOS}. Usually (and especially for
	equal or nearly equal mergers that do not promptly collapse to a BH), the
	synthesis of light $r$-process elements within BNS ejecta should be favoured by
	soft \acp{EOS}, since the higher temperatures achieved in the shock-heated
	ejecta component leptonise matter in a more efficient way. However, we notice
	that for $A \lesssim 120$ the relative production of light $r$-process elements
	does not follow exactly this trend. This is because, for such asymmetric
	binaries promptly collapsing to BHs, the dynamical ejection of matter is usually
	dominated by the cold, neutron-rich tidal component. However a small, but
	non-negligible fraction of the dynamical ejecta comes from the contact surface
	of the colliding NSs and is characterised by relatively high entropies (see the
	$X_{\rm s}$ column in \reftab{tab:ejecta}). The corresponding larger peak
	temperatures produce a tail in the $Y_e$ distribution above $\approx 0.22$.
	These ejecta are likely present in all BNS mergers, but their relatively low
	amount make them more relevant only in the case of mergers characterised by a
	very small dynamical ejecta mass. Moreover, these ejecta can more likely escape
	in the case of stiffer \acp{EOS}, characterised by larger radii and less deep
	gravitational well.
	
	We conclude that the nucleosynthesis patterns show a mild variability, depending
	on the mass ratios and \acp{EOS}. However, they are comparable with the ones
	obtained by BNS merger simulations of lighter binary systems and do not show
	peculiar behaviours \citep[see
	e.g.][]{Wanajo:2014wha,Just:2014fka,Radice:2018pdn,Bovard:2017mvn,Nedora:2020pak}.
	Nevertheless, we point out that the nucleosynthesis yields obtained exhibit
	different features with respect to the Solar residuals, for example in the
	position and shape of the second and third $r$-process peaks. The fine structure
	of the abundance pattern in this region is indeed affected by the particular
	choice of the nuclear input data made for the nucleosynthesis calculations, like
	for example the nuclear mass model, the different fission channels considered
	(spontaneous, neutron-induced, $\beta$-delayed etc.) or the fission fragment
	distribution employed \citep[see
	e.g.][]{Eichler:2014kma,Mendoza-Temis:2014mja,Goriely:2015oha}.
	However, since we do not expect dynamical ejecta from high-mass \ac{BNS} mergers to
	represent the dominant contribution to the $r$-process enrichment in the Universe, possible discrepancies with the solar pattern are not an issue. In addition,
	one should also remember that, even for high mass \ac{BNS} mergers, the nucleosynthesis from the disc ejecta is expected to dominate the dynamical ejecta one.
	
	
	\subsection{Kilonovae}
	\label{sec:kilonovae}
	\begin{figure*}
		\includegraphics{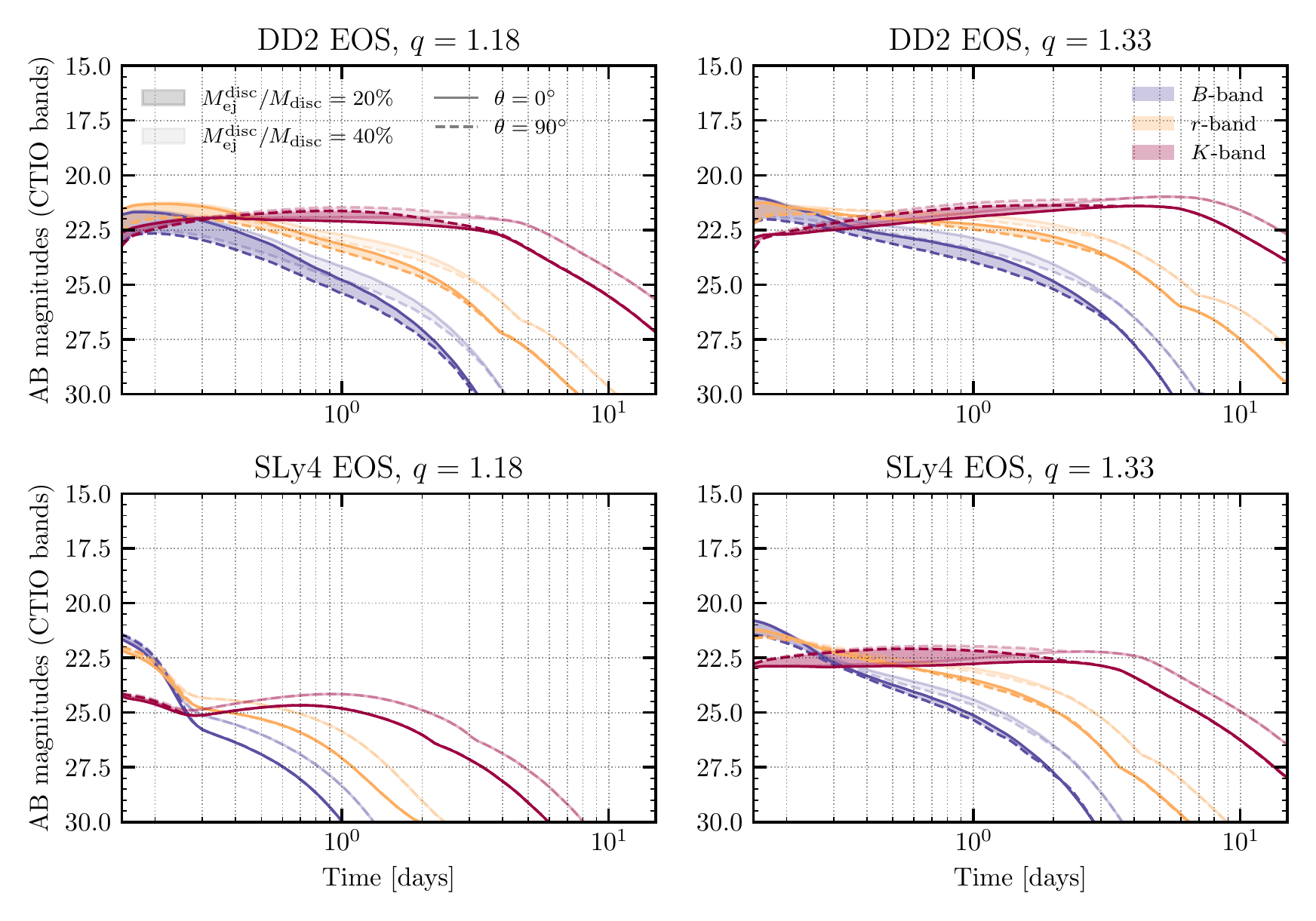}
		\caption{AB magnitudes in the blue, red and IR bands of CTIO telescope as a
			function of time. We report the results for the DD2 and SLy4 \acp{EOS} and for
			a binary mass ratio of $q=1.18$ and $q=1.33$ at standard resolution. The
			uncertainty in the source inclination angle (varying between
			$0^\circ-90^\circ$) is represented using solid lines for $\theta=0^\circ$ and
			dotted lines for $\theta=90^\circ$, with intermediate values enclosed by the
			above lines. The source distance is set to $130~{\rm Mpc}$.In each panel, the darker and lighter areas refer to two different scenarios in which $20\%$ and $40\%$ of the disc mass is expelled, respectively.}
		\label{fig:mag_2x2}
	\end{figure*}
	
	Using the model described in \refsec{subsec:kilonova calculations}, we compute
	synthetic kilonova light curves for each of the SR models presented in this work
	for which the mass of the dynamical ejecta is larger than $10^{-5} \Msun$. In
	\reffig{fig:mag_2x2}, we present the evolution of the AB magnitudes in three
	representative bands ($B$-, $r$-, and $K$-band), for two \acp{EOS} (the stiff
	DD2 and the soft SLy4) and two mass ratios ($q=1.18$ and $q=1.33$). In general,
	kilonova magnitudes depend both on the distance and on the viewing angle.
	Regarding the former, the wide range of distances compatible with GW190425
	($D=70-250$~Mpc) implies a possible uncertainty of $\sim 3$ magnitudes, with
	lower magnitudes corresponding to shorter distances. On the other hand, the
	inclination angle is almost unconstrained by the GW190425 signal. Due to the
	degeneracy between viewing angle and distance, viewing angles close to the polar
	axis ($\theta_{\rm view} \sim 0^\circ$) are more compatible with larger
	distances, while shorter distances would imply edge-on configurations
	($\theta_{\rm view} \sim 90^\circ$). In \reffig{fig:mag_2x2}, we set $D=130~{\rm
		Mpc}$ while we explore all possible viewing angles, $\theta_{\rm view} \in [
	0^\circ,90^\circ ]$.
	The amount of ejecta and their composition are the most relevant parameters in
	shaping kilonova light curves. In general, since GW190425-like events are
	expected to eject a relatively small amount of mass, the resulting kilonovae are
	predicted to be relatively dim and fast-evolving, compared for example with
	GW170817-like events.
	More specifically, in \reffig{fig:mag_2x2} we observe that the kilonova
	associated to the simulation employing the DD2 \ac{EOS} and with $q=1.33$ is
	brighter and lasts longer with respect to both the simulation employing the same
	\ac{EOS} but with $q=1.18$, and the simulation with the same mass ratio but
	employing the SLy4 \ac{EOS}, for all bands. This mostly reflects the difference
	in the amount of ejecta between the different models, see
	\refsec{sec:remnant_properties} and \refsec{sec:dynamical_ejecta}, with greater
	mass ejection resulting in brighter peak luminosities due to the stronger
	availability of nuclear fuel required for the kilonova emission.
	
	Differences in the viewing angle affect the light curves at times shorter than a
	couple of days, while our results are insensitive to the specific viewing angle
	at later times. This can be explained by considering that the slower and
	significantly more massive disc wind component, eventually powering the kilonova
	at late times ($t \gtrsim 1$ day), is assumed to be isotropic in our model.
	Conversely, within the first days after merger, the dynamical ejecta component
	plays a relevant role. The angular distribution of its mass and composition are
	thus reflected in the band magnitude evolution. In particular, we obtain
	brighter light curves in the visual bands at angles closer to the pole
	($\theta\sim0\degr$), where matter with a higher initial $Y_e$ (and thus lower
	opacity) can be found. Conversely, the emission
	in the IR band is typically brighter close to the equatorial plane ($\theta \sim
	90^\circ$), where the most neutron-rich (and thus more opaque) matter is
	concentrated, with respect to higher latitudes. Since for each of our SR models
	the disc wind ejecta component is determinant in generating the kilonova
	emission, we test our results sensitivity with respect to its mass.
	In particular, we notice that the increase in the fraction of ejected
	disc mass from a plausible $20\%$ to an optimistic $40\%$ results in an overall
	gain in brightness of $\sim1$ magnitude for all bands at late times, when the
	disc ejecta component becomes dominant. We also test the sensitivity of light
	curves on the disc ejecta mass and composition angular distributions. We consider
	a density distribution $\rho_{\rm wind}(\theta)\propto \sin{\theta}$ as alternative
	to the isotropic case and an opacity distribution shaped as a step function
	with $k=1$ ${\rm cm^2~g^{-1}}$ for $\theta<45^\circ$ and $k=10$ ${\rm cm^2~g^{-1}}$
	for $\theta>45^\circ$. While such modifications on the opacity can vary the final
	bolometric light curves up to a factor of a few, the different mass
	distribution results in a model dependence on the viewing angle also at late
	times. More specifically, since the wind density gradually increases towards the
	equator, the magnitudes decrease accordingly for all bands, and we obtain the
	brightest emission for $\theta_{\rm view}\sim90^\circ$, $\sim1$ magnitude below
	the polar one. Despite the non-negligible dependences, these tests place
	our uncertainty in the luminosity due to the disc parameters well below
	the one due to the source distance and viewing angle.
	
	For simulations with $q=1.33$, providing a prominent tidal low-$Y_e$ ejecta
	component, the infrared $K$-band lasts several days and nearly always dominates
	over bluer bands, due to the prevailing presence of lanthanides-rich material
	synthesised through a strong $r$-process both in the dynamical and in the disc
	wind ejecta. On the other hand, in the case of the simulation with $q=1.18$ and
	the SLy4 \ac{EOS}, the considerably lower ejecta mass with a broader $Y_e$
	distribution results in lower material opacities and slightly brighter blue band
	light curves at early times.
	
	\begin{figure*}
		\includegraphics{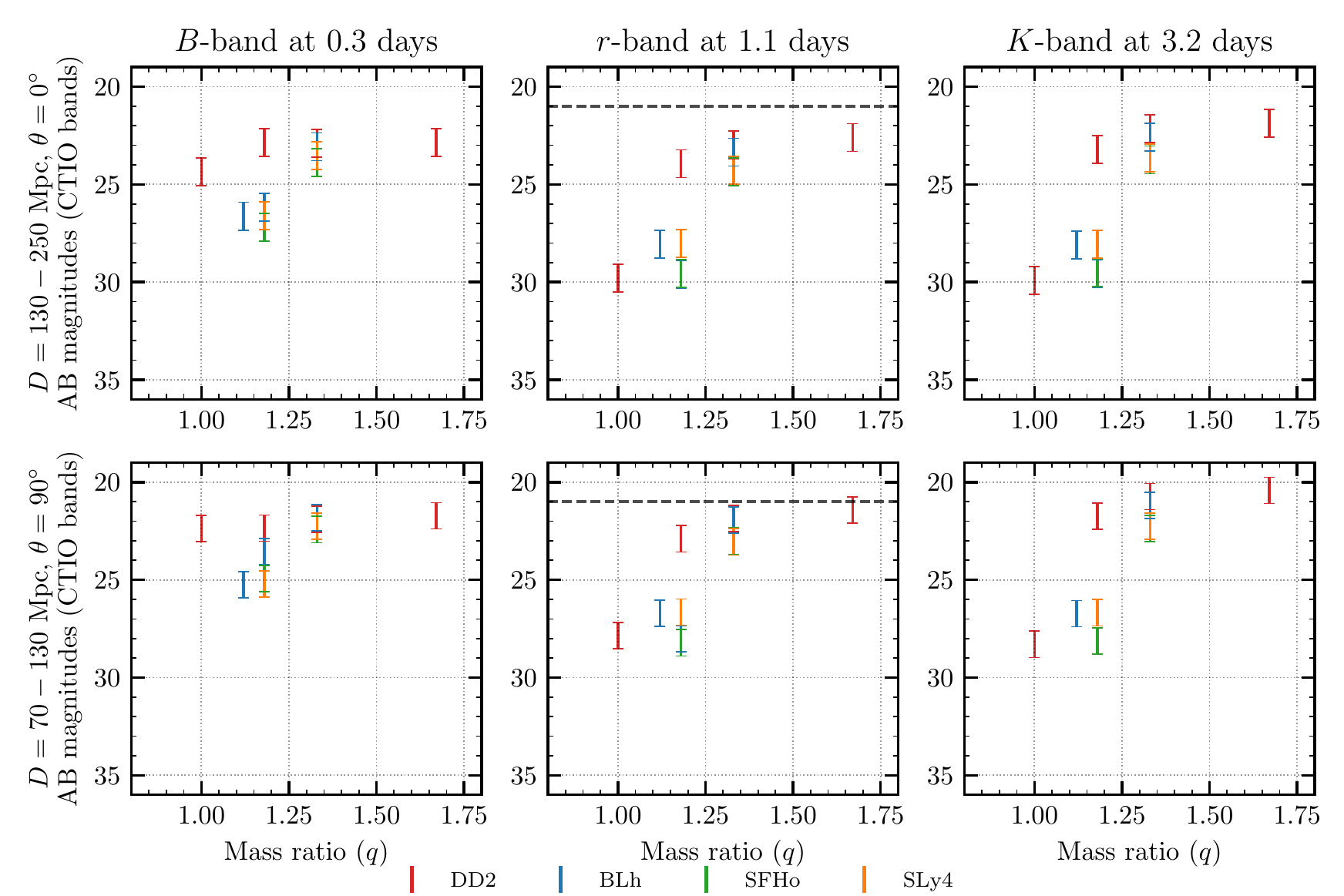}
		\caption{AB magnitudes in the blue, red and IR bands of CTIO telescope at fixed
			characteristic times as a function of the binary mass ratio $q$. The kilonova
			is obtained assuming an ejection of $20\%$ of the disc mass. Results are
			colour-coded to indicate different \acp{EOS}. Only standard resolution
			simulations are shown. Two cases for the source distance and inclination angle
			are reported, with the error bars representing the uncertainty in the source
			distance. The dashed horizontal line represents the upper limit for GW190425
			obtained with the ZTF by the GROWTH collaboration for the $r$ and $g$-band
			\citep{Coughlin:2019xfb}.}
		\label{fig:mag_3x2}`
	\end{figure*}
	Due to the evolution of the photospheric temperature, the $B$-band magnitude is
	the first to peak, within the very first few hours, promptly followed by the
	$r$-band magnitude, dominating within the first half-day after merger, while the
	infrared band peaks much later in time, possibly on a time-scale of days. While
	the precise peak times and magnitudes vary depending on the specific simulation,
	the presence of common trends in the light curve behaviour allow us to identify
	characteristic time-scales for each band in which the latter typically dominates
	over or is comparable to the others. In \reffig{fig:mag_3x2}, we present the
	values of the AB magnitudes in the same three bands as in \reffig{fig:mag_2x2}
	at three corresponding characteristic times for each available simulation,
	namely at 0.3 days, 1.1 days and 3.2 days for the $B$, $r$ and $K$ band,
	respectively. Since we want now to address the possible detectability of
	GW190425, two possible ranges for the source distance and inclination angle are
	considered in order to account for the large degeneracy in the estimation of
	these parameters for GW190425 \citep[see also ][for a similar
	choice]{Dudi:2021abi}. Regardless of the specific band, magnitudes tend to
	decrease with the increase of the mass ratio, leading to emissions up to $\sim8$
	magnitudes brighter, moving from equal-mass to strongly asymmetric mergers.
	Likewise, the stiffest \ac{EOS} corresponds to luminosities which can be as
	bright as $\sim6$ magnitudes below the same results obtained using softer
	\acp{EOS}. Exceptions to these trends can be directly traced back to already
	emerged distinctive mass ejections. For example, the simulation employing the
	BLh \ac{EOS} and a mass ratio of $q=1.12$ returns brighter red and infrared
	luminosities with respect to the simulation employing the same \ac{EOS} but with
	$q=1.18$: this is due to the fact that in the first instance the computed disc
	mass is greater, leading to a more massive disc wind (which dominates over the
	dynamical component). Based on our analysis, from \reffig{fig:mag_3x2} it is
	clear that almost none of our models can be fully ruled out by the ZTF upper
	limits to the kilonova of GW190425 (shown as a dashed horizontal line), meaning
	that current data cannot help further constraining the model parameters. This
	leaves open the question as to whether the detection of events like GW190425 can
	shed light on the source properties, and hints to the necessity of determining
	the sky localisation with high accuracy for these events, to employ deeper
	observations in order to resolve such EM counterparts.
	
	\section{Discussion}
	\label{sec:discussion}
	
	In this section, we compare the results of our work with recent publications
	about the modelling of GW190425 and of its EM counterparts, in particular with
	results reported in \citet{Dudi:2021abi,Raaijmakers:2021slr,Barbieri.etal:2021}.
	
	During the preparation of this work, \Dudi{} published an independent study on
	GW190425 in \ac{NR}. They used the \texttt{BAM} code, a \ac{NR} code which was
	shown to produce results consistent with \texttt{WhiskyTHC} \citep[see
	e.g.][]{Dietrich:2018phi}. They considered four mass ratios, ranging from 1 to
	1.43, and for each of them they employed three cold, beta-equilibrated
	\acp{EOS}: the piecewise-polytropic \ac{EOS} MPA1 \citep{Read:2008iy}, a
	piecewise-polytropic representation of the tabulated DD2 \ac{EOS} at the lowest
	available temperature, and the softer APR4 \ac{EOS} \citep{Akmal:1998cf}. Each
	model was run at three different resolutions, with our SR being intermediate
	between their worst and middle resolution.
	Similarly to what we found in our simulations, all the BNS models presented by
	\Dudi{} result in a prompt collapse. Regarding the properties of the remnant,
	the two works predict a comparable range for $M_{\rm BH}/M$, while we notice
	that the dimensionless spin parameter obtained by \Dudi{} is systematically
	lower than the one obtained by our simulations by several percents,
	corresponding to $\Delta a_{\rm BH} \sim 0.05$, when comparing simulations
	characterised by similar mass ratios and \acp{EOS}. Both analyses agree in
	predicting more massive discs when considering more asymmetric binaries and
	stiffer \acp{EOS}. In particular, the disc results for the DD2 \ac{EOS} share
	the same trend with respect to $q$, both on a qualitative and quantitative
	level.
	Moving to the comparison of the dynamical ejecta, we first notice that the
	amount of matter obtained for the MPA1 and APR4 \acp{EOS} by \Dudi{} increases
	as the binary becomes more asymmetric, similarly to what observed in our BLh,
	SFHo and SLy4 simulations. Similarly, the amount of ejecta from the DD2
	simulations first increases then decreases with $q$ in both analyses. However,
	while in the former cases the amount of ejecta are comparable among them, the
	values obtained for the DD2 \ac{EOS} differ significantly, with the ejecta
	reported in \Dudi{} larger by $\sim$ one order of magnitude. According to the
	reported values, uncertainties due to different resolutions seem to account only
	for a fraction of this discrepancy and higher resolution seems to result in
	smaller ejecta masses. A potentially relevant source of discrepancy could be the
	different microphysical input. In addition to a more accurate temperature
	treatment, the presence of neutrino radiation can influence the dynamical
	ejecta, since simulations accounting for neutrino emission show systematically
	smaller dynamical ejecta masses \citep[see e.g.][]{Nedora:2020qtd}, due to the
	emission of neutrinos occurring during the ejection process.
	
	The different amount of ejecta obtained employing the DD2 \ac{EOS} is directly
	reflected in the kilonova light curves, where for a similar mass ratio the
	$r$-band magnitudes reported in \Dudi{} are systematically brighter. In
	particular, while for edge-on views the results are in good agreement, for a
	viewing angle close to the polar axis we find up to $\sim5$ magnitudes of
	difference between light curves corresponding to the same binary configurations.
	On the one hand, this may reflect the substantially different mass and
	composition distributions resulting from the \ac{NR} models. On the other hand, we
	also stress that the models employed for the light curves computation are
	significantly different: as opposed to our semi-analytic model described in
	\refsec{subsec:kilonova calculations}, \Dudi{} employ a more advanced
	wavelength-dependent radiative transfer approach \citep{Kawaguchi:2019nju}, for
	which the post-merger ejecta composition is fixed for all components.
	Additionally, our kilonova model decomposes the solid angle in radial slices.
	While this approach is reasonable for ejecta expelled over the entire solid
	angle, it could be inadequate for ejecta expelled only close to the equator for
	which it tends to underestimate magnitudes up to a few since it neglects
	possible lateral effects
	\citep{Kawaguchi:2016ana,Kawaguchi:2018ptg,Barbieri:2019sjc, Bernuzzi:2020txg}.
	Keeping in mind the above differences for the GW190425 event and working under
	the assumption that the location of the source was covered by ZTF, \Dudi{}
	disfavored a higher number of models with respect to this work, i.e., the ones
	employing DD2 or MPA1 \acp{EOS} with a high mass ratio and a source
	configuration similar to that used in the top panels of \reffig{fig:mag_3x2}. On
	the contrary, our results imply that only the model employing the DD2 \ac{EOS}
	with the highest mass ratio and a source distance close to $D\sim70$~Mpc
	(corresponding to a edge-on view) would be disfavoured (as visible in the bottom
	panels of \reffig{fig:mag_3x2}).

	\citet{Raaijmakers:2021slr} studied the expected photometric light curves of BNS
	mergers with masses in the range compatible with the posteriors of GW190425. We
	recall that, due to the spherical symmetry of the employed kilonova model, it
	was not possible to investigate the light curve dependence on the viewing angle,
	even if selected tests with the multidimensional \texttt{POSSIS} code were
	performed \citep{Bulla:2019muo}. By fixing the source distance to $130$~Mpc, we
	find that the spread in the magnitudes generated by the different \ac{NR} models
	considered in this work is comparable to the comprehensive results displayed in
	\citet{Raaijmakers:2021slr}, which span $\sim4$ magnitudes at times shorter than
	$\sim1$ day. In the same time period, our light curves are generally dimmer with
	respect to those computed in \citet{Raaijmakers:2021slr}, with an average
	difference of $\sim3$ magnitudes. A plausible source of this systematic
	discrepancy lies in the different ways in which the ejecta and disc masses were
	computed. In our case, they are the outcome of BNS merger simulations, while in
	\citet{Raaijmakers:2021slr} they are estimated on the basis of the fitting
	formulae for the mass of the dynamical ejecta and of the disc proposed in
	\citet[equations 4 and 6]{Kruger:2020gig}, and for the average dynamical ejecta
	speed proposed in \citet{Foucart:2016vxd}. These formulae take as input
	parameters the compactness and the masses of the binary components. We compare
	the outcome of these fitting formulae with our numerical results in
	\refapp{sec:fitting_formulae}. We found significant differences in the ejected mass and in the expansion speed, and less severe disagreement for the disc mass, which is consistent with the numerical data when errors are taken in consideration. In particular, the mass
	of the ejecta predicted by the fitting formulae is $\sim 10-100$ higher than in
	our simulations. Our comparison reveals how \ac{NR} fitting formulae can become inaccurate when used far from their calibration regime.
	
	Finally, we compare the light curves computed in this work with those obtained
	in \citet{Barbieri.etal:2021} for BNS systems, and, as in the case of
	\citet{Raaijmakers:2021slr}, we find typically lower peak luminosities. Since
	also \citet{Barbieri.etal:2021} used fitting formulae to predict the ejecta
	properties (see
	\refapp{sec:fitting_formulae} for a more detailed discussion), we argue that disc and ejecta masses larger by one or even two
	orders of magnitudes can account for the observed differences. In addition, our
	results employing the DD2 \ac{EOS} are significantly more sensitive to the binary
	configuration, as peak luminosities in the $r$-band and at IR frequencies vary
	by $\lesssim 7$ magnitudes for a mass ratio varying between $1\leq
	q\lesssim1.7$, while in \citet{Barbieri.etal:2021} the same bands exhibit a
	variation of $\sim3.5$ magnitudes for a mass ratio between $1\lesssim
	q\lesssim2$. Also in this case, at least a part of these differences is possibly
	due to disc later irradiation, which is expected to occur in very asymmetric
	system, which was taken into account by \citet{Barbieri.etal:2021}.
	
	Both in \citet{Raaijmakers:2021slr} and \citet{Barbieri.etal:2021}, the overall
	brighter kilonovae allow the identification of some binary configurations
	potentially detectable by the ZTF within the first few days from merger, in
	addition to a major portion of the BHNS configurations considered in those
	works. In particular, in \citet{Barbieri.etal:2021} several configurations
	employing the DD2 \ac{EOS} and the APR4 \ac{EOS} can be ruled out by the GW190425 EM
	follow-up. Conversely, here almost all the our BNS simulations employing the DD2
	\ac{EOS} and the totality of those employing softer \acp{EOS} produce kilonovae which are
	not detectable by ZTF in a GW190425-like event at a comparable distance.
	\section{Conclusions}\label{sec:conclusions}
	
	In this work, we investigated in detail the outcome of BNS merger simulations
	targeted to GW190425 with detailed microphysics. We set up 28 simulations with
	finite temperature, composition dependent NS \acp{EOS}, and neutrino radiation. For
	each simulation we extracted remnant and dynamical ejecta properties, and we
	computed in post-processing nucleosynthesis yields and kilonova light curves.
	Using 4 \acp{EOS} compatible with present constraints and considering a broad range
	of mass ratios, we aimed at giving an accurate description of GW190425-like BNS
	mergers and answering a number of questions, including: what can we expect from
	future detection of this kind of events in terms of remnant, dynamical ejecta,
	nucleosynthesis signature and kilonova light curves? Despite the wide sky
	localisation of GW190425, can the lack of an EM counterpart give constraints on
	the \ac{EOS} and/or the binary parameters?
	
	We found that such BNS mergers, characterised by an unusual high total mass of
	$3.4~\Msun$ and a chirp mass of $1.44 \Msun$, prompt collapse to a light black
	hole of $\sim 3.2~\Msun$ with a dimensionless spin parameter that ranges from
	0.73 to 0.83, surrounded by a light disc formed by tidal interactions.
	Asymmetric BNS mergers with stiffer \ac{EOS} have more massive remnant disc, ranging
	from $10^{-5}~\Msun$ for equal mass binaries with soft \ac{EOS}, to $0.1~\Msun$ for
	the most asymmetric BNS in our sample.
	
	During the late inspiral and merger, previous to the collapse, the simulated
	binaries expel a small amount of matter in the form of dynamical ejecta. The
	high compactness is responsible for less deformable NSs while the prompt
	collapse inhibits the production of shock-heated ejecta. This explains the lower
	values of ejected mass compared to what previously found for BNS whose chirp
	mass is closer to what is observed in the Galactic BNS population and in
	GW170817. Since tidal interactions are the main cause of dynamical ejection, we
	found that asymmetric BNS mergers with a stiff \ac{EOS} are able to unbind up to
	$\sim 10^{-3}~\Msun$ of ejecta, while equal mass BNS with a soft \ac{EOS} only eject
	$\lesssim 5 \times 10^{-6}~\Msun$ of matter. Also the properties mostly depend
	on the mass ratio and on the \ac{EOS} of the BNS merger. Dynamical ejecta spread all
	over the space but it is mainly concentrated along the orbital plane in an
	opening angle which goes from $54^\circ$ for symmetric BNS to $18^\circ$ for the
	more asymmetric BNS in our sample. We also discuss the distributions of electron
	fraction, velocity at infinity and entropy of the dynamical ejecta and their
	trends with the binary parameters.
	
	In all the considered simulations, the resulting $r$-process nucleosynthesis
	pattern does not show peculiar behaviours and reflects directly the properties
	of the matter outflow. For ejecta dominated by cold, neutron-rich matter, we
	noticed a remarkably robust production of heavy elements between the second and
	the third $r$-process peaks, as opposed to the less significant one of lighter
	elements. The latter is however more sensitive to the binary parameters. In
	fact, around the first peak the nucleosynthesis pattern changes depending on the
	\ac{EOS} considered (even if not with a clear trend) and increases with decreasing
	mass ratio, but always on a lower level with respect to the Solar residuals.
	
	For the kilonova, we found that narrow-band light curves in the $B$- and $r$-
	bands peak within the first few hours after the merger with a rapid subsequent
	decline, while the emission at IR frequencies lasts several days. Assuming a
	distance of 70-130 Mpc or 130-250 Mpc, compatible with what was inferred for
	GW190425, and combined with a edge-on or face-on inclination, respectively, the
	peak magnitude in every band is not brighter than $\sim20$ magnitudes, as
	opposed to the case of kilonovae resulting from BNS more compatible with the
	Galactic BNS population or with GW170817. As such, we conclude that it could be
	difficult to observe such a transient at the distances inferred for GW190425
	with present wide-field surveys, unless a good sky localisation allows for
	deeper and localised searches. This can be traced back to the low mass of the
	dynamical ejecta and of the disc remnant. Only a BNS with a particularly stiff
	\ac{EOS}, a high mass ratio and a source distance around $\sim70$~Mpc would have been
	detected by the ZTF facility according to our findings. This would favour a
	BH-NS merger in the case of a kilonova detection resulting from a compact binary
	merger similar to GW190425 by ZTF.
	
	Future follow-up campaigns will be joined by Vera Rubin (LSST) observatory. In
	spite of the relatively small field of view ($\sim 10~{\rm deg}^2$) compared to
	ZTF, the short read-out time, the all-sky reference and a sensitivity of
	$24.7-27.5$ AB magnitudes in the $r$-band will permit Vera Rubin to be a
	powerful resource to detect faint kilonovae \citep{Andreoni:2021epw}. Vera Rubin
	is potentially able to detect kilonova signals from some of the simulated BNS
	mergers. For a kilonova at a distance of $130-250~{\rm Mpc}$, a kilonova signal
	would be detectable for BNS mergers with $q>1.33$ and, in the case of a very
	stiff \ac{EOS} (as DD2) for the BNS with $q=1.18$. In addition, for smaller
	distances, i.e. $70-130{\rm Mpc}$, also kilonovae resulting from slightly
	asymmetric BNS mergers could be observable. Finally, for a distance comparable
	to the one of GW170817, all the simulated kilonovae could be potentially
	detected. However, despite the increased sensitivity, Vera Rubin's field of view
	will cover efficiently up to $200~{\rm deg}^2$, far less than the confidence
	region of GW190425. Thus, a better sky localisation will be crucial.
	
	We compared our results with recent works that aim to predict the remnant and
	ejecta properties, as well as the kilonova light curves of GW190425. We find
	overall similar qualitative trends, but with some quantitative differences. In
	the case of \Dudi, who explored a comparable set of simulations in numerical
	relativity, trends in the ejecta masses and disc masses are very similar, with a
	better quantitative agreement for the latter than for the former. We speculate
	that these differences could be due to the different microphysical setups (both
	polytropic \acp{EOS} and the lack of neutrino radiation tend to overestimate the
	dynamical ejecta) as well as resolution effects. All these uncertainties could
	be even amplified in this case due to the small amount of ejecta, that makes
	their identification and tracking inside the computational domain more
	challenging. \citet{Raaijmakers:2021slr} and \citet{Barbieri.etal:2021} computed
	kilonova light curves for GW190425-like events and they found kilonova
	transients systematically brighter than ours. A plausible source of discrepancy
	could be the use of existing fitting formulae to predict the dynamical ejecta
	and the disc mass. Indeed the peculiarity of GW190425 slip to the predictions
	given by the formulae presented in previous works
	\citep{Foucart:2016vxd,Nedora:2020qtd,Barbieri:2019sjc,Radice:2018pdn} that we
	took into exam. Fitted on large sample of numerical simulations of BNS mergers
	with parameters however different from the ones of GW190425, they usually
	predict an enhancement of the dynamical ejecta and of the disc mass with respect
	to our simulations, with observable consequences on the kilonova. This result
	underlines the difficulty in providing fitting formulae for the ejecta
	properties valid over a broad range of binary parameters and even outside of the
	fitting range. This could indeed strongly affect their effectiveness.
	
	The detection of GW190425 demonstrated that, in addition to the sample of BNS
	mergers whose properties are close to the ones observed in the current
	population of Galactic BNS systems, there could be a population of GW-loud
	events characterised by larger chirp masses. Their modelling is less developed
	and their properties (including the smaller ejecta and disc masses) are possibly
	more challenging to study. Our work represents a step forward in the direction
	of better characterising such systems. Considering the GW190425 follow-up
	campaign, we conclude that, even assuming that the sky coverage was enough and
	the binary was a BNS system, no strong constraints on the BNS parameters nor on
	the \ac{EOS} can be inferred by the lack of EM signal. Only the corner case of very
	stiff \ac{EOS} and extreme mass ratios could be possibly excluded. Future
	observations of EM counterparts by wide-field surveys, such as ZTF or Paolmar
	Gattini-IR telescope, for such a population outsider will be non trivial, unless
	the merger distance decreases to $\lesssim40$ Mpc.
	However, large uncertainties still remain. We mostly quantified errors due to
	finite resolutions, but we expect possibly larger uncertainties due to
	systematics and modelling limitations. Further works in the modelling of both
	BNS mergers and their EM counterparts is required to properly assess these
	limitations.
	
	\section*{Acknowledgements.} 
	We thank Andrea Endrizzi for initial work on the project. The Authors
	acknowledge the INFN and Virgo for the usage of computing and storage resources
	through the \texttt{tullio} cluster in Torino. AP acknowledge PRACE for awarding
	him access to Joliot-Curie at GENCI@CEA. He also acknowledges the usage of
	computer resources under a CINECA-INFN agreement (allocation INF20\_teongrav and
	INF21\_teongrav). S.B.~acknowledges funding from the EU H2020 under ERC Starting
	Grant, no.BinGraSp-714626, and from the Deutsche Forschungsgemeinschaft, DFG,
	project MEMI number BE 6301/2-1. D.R.~acknowledges funding from the U.S.
	Department of Energy, Office of Science, Division of Nuclear Physics under Award
	Number(s) DE-SC0021177 and from the National Science Foundation under Grants No.
	PHY-2011725, PHY-2020275, PHY-2116686, and AST-2108467. FMG acknowledges funding
	from the Fondazione CARITRO, program Bando post-doc 2021, project number 11745.
	\ac{NR} simulations were performed on Joliot-Curie at GENCI@CEA (PRACE-ra5202),
	SuperMUC-LRZ (Gauss project pn56zo), Marconi-CINECA (ISCRA-B project HP10BMHFQQ,
	INF20\_teongrav and INF21\_teongrav allocation); Bridges, Comet, Stampede2 (NSF
	XSEDE allocation TG-PHY160025), NSF/NCSA Blue Waters (NSF AWD-1811236),
	supercomputers. This research used resources of the National Energy Research
	Scientific Computing Center, a DOE Office of Science User Facility supported by
	the Office of Science of the U.S.~Department of Energy under Contract
	No.~DE-AC02-05CH11231. 
	
	\section*{Data availability}
	Data generated for this study will be made available upon reasonable request to
	the corresponding authors.
	
	\bibliographystyle{mnras}
	\bibliography{refs,local}
	
	\appendix\label{sec:appendix}

	\section{Details of the keplerian model}\label{app:keplerian}
	To deduce \refeq{eq:keplerian_ratio} we define
	\begin{align}
	M_{\rm disc}^{\rm kep} \equiv M_{\rm disc}^{\rm G} + M_{\rm disc}^{\alpha}~, &&
	J_{\rm disc}^{\rm kep} \equiv J_{\rm disc}^{\rm G} + J_{\rm disc}^{\alpha}~,
	\end{align}
	where the superscript G and $\alpha$ indicate the Gaussian and power-law parts
	of the Keplerian disc in \refeq{eq:disc_radial_density} and
	\refeq{eq:mass_J_keplerian}:
	\begin{align}
	M_{\rm disc}^{\rm G} \equiv \int_{R_{\rm ISCO}}^{r^*} r \sigma(r) {\rm d}r~, &&
	M_{\rm disc}^{\rm \alpha} \equiv \int_{r^*}^{r_{\rm max}} r \sigma(r) {\rm d}r~,
	\end{align}
	and similar for the angular momentum. We can solve the integration:
	\begin{align}
	M_{\rm disc}^{\rm G} &= bs^2
	\left[ 
	\sqrt{\frac{\pi}{2}} \frac{r_{\rm peak}}{s} {\rm erf}\left( \frac{r - r_{\rm peak}}{\sqrt{2} s} \right) - 
	\exp \left(- \frac{(r - r_{\rm peak})^2}{2 s^2} \right)
	\right]_{R_{\rm ISCO}}^{r*}~, \\
	M_{\rm disc}^{\alpha} &= \frac{\sigma_0}{\alpha -2 } \left( 
	1 - \frac{(r^*)^{\alpha - 2}}{r_{\rm max}^{\alpha - 2}}
	\right)(r^*)^2~,
	\end{align}
	\begin{align}
	J_{\rm disc}^{\rm G} &= \sqrt{\frac 1 2 G M_{\rm BH} r_{\rm peak}^3 (bs)^2} \sum_{k=0}^{\infty} \binom{3/2}{k} \left( \frac{ \sqrt{2} s}{r_{\rm peak}} \right)^k \\
	&~\times \left. \Gamma \left( \frac{k+1}{2},~\frac{(r - r_{\rm peak})^2}{2 s^2} \right) \right|_{R_{\rm ISCO}}^{r^*}~,\\
	J_{\rm disc}^{\alpha} &= \frac{\sigma_0 \sqrt{G M_{\rm BH}}}{\alpha - 5/2} \left( 1 -
	\frac{(r^*)^{\alpha - 5/2}}{r_{\rm max}^{\alpha - 5/2}}
	\right)(r^*)^{5/2}~,
	\end{align}
	where ${\rm erf}(x) \equiv (2/\sqrt{\pi}) \int_0^x e^{-t}~{\rm d}t$ is the error
	function and $\Gamma(a,x) \equiv \int_{x}^{\infty} t^{a-1} e^{-t} dt~$ the
	incomplete gamma function.
	One can write:
	\begin{equation}
	\frac{J^{\rm kep}_{\rm disc}}{M^{\rm kep}_{\rm disc}} =
	\eta \frac{J^{\alpha}_{\rm disc}}{M^{\alpha}_{\rm disc}}~,
	\end{equation}
	where 
	\begin{equation}\label{eq:eta}
	\eta = \frac{1 + J^{\rm G}_{\rm disc} / J^{\alpha}_{\rm disc}}{1 + M^{\rm G}_{\rm disc} / M^{\alpha}_{\rm disc}} ~.
	\end{equation}
	Assuming $r^* \ll r^{\rm max}$ (with an error $\lesssim 1$ per cent) we arrive at
	\begin{equation}
	\frac{J^{\rm kep}_{\rm disc}}{M^{\rm kep}_{\rm disc}} = \eta \frac{\alpha - 2}{\alpha - 5/2} \sqrt{G M_{\rm BH} r^*}~.
	\end{equation}
	As showed in \reffig{fig:mass_ang_fit}, the model tends to underestimate the
	radial angular momentum density, especially for $r < r^*$.
	\begin{figure}
		\includegraphics{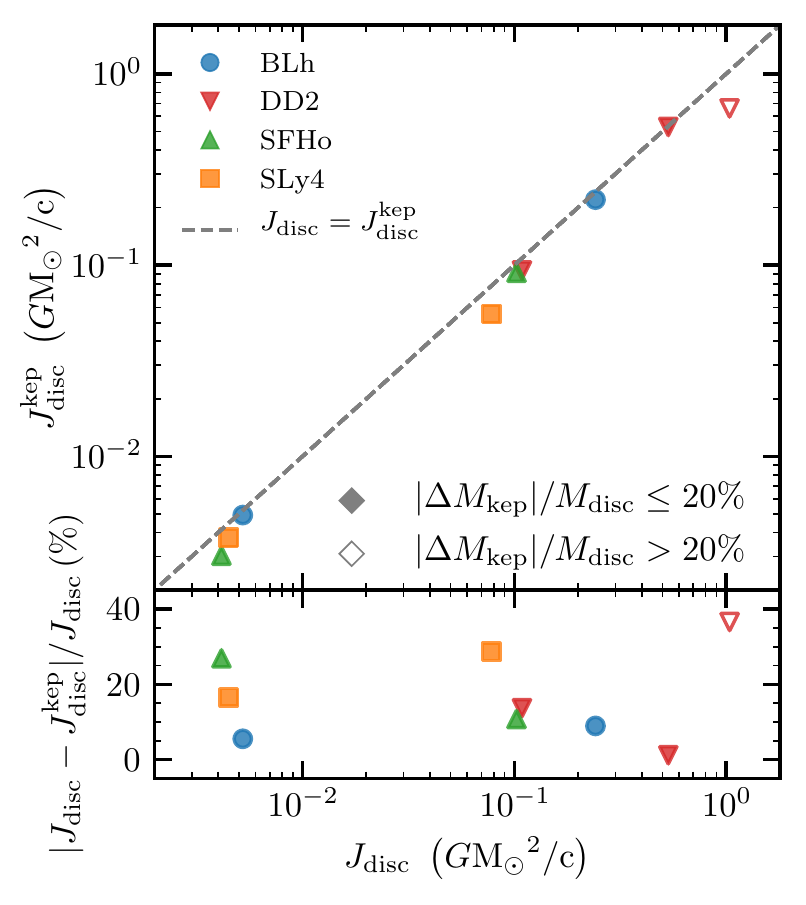}
		\caption{Top: Comparison between the disc angular momentum outside the ISCO
			from numerical simulations, $J_{\rm disc}$, and the one obtained by
			constructing a Keplerian disc whose radial density profile was fitted
			over the numerical results using \refeq{eq:disc_radial_density}, $J_{\rm
				disc}^{\rm kep}$. Bottom: Relative difference between the two values.
			Unfilled markers represent discs for which the Keplerian mass differs from
			the numerical one by more than 20 per cent.}
		\label{fig:keplerian_disc}
	\end{figure}
	To better quantify this difference, in \reffig{fig:keplerian_disc} we compare
	the angular momentum of the discs from our simulations at SR with the
	corresponding Keplerian analogue,\refeq{eq:mass_J_keplerian}. With the exception
	of DD2 EOS with $q=1.67$, the discrepancy is $<$30 per cent.
	%
	\begin{figure}
		\includegraphics{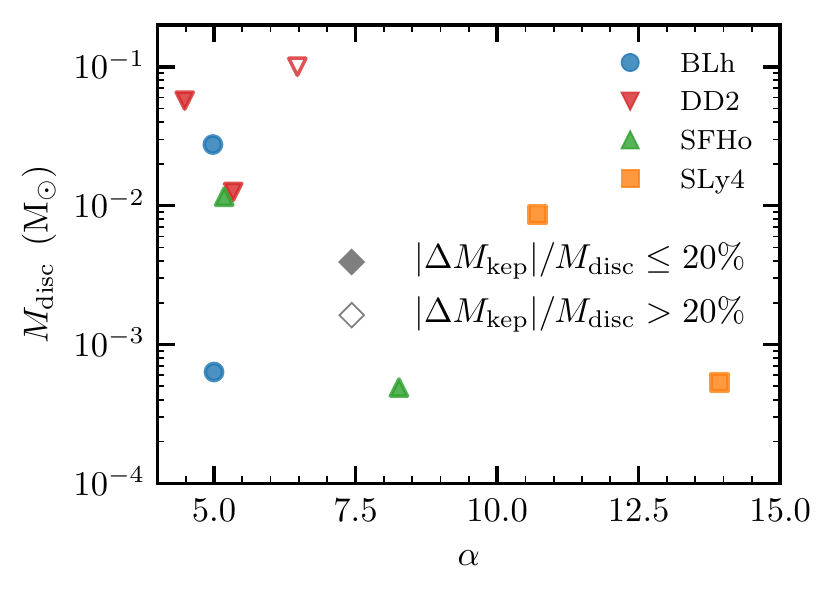}
		\caption{Power-law exponent, $\alpha$, for each disc in our numerical
			simulation sample, as a function of the disc mass, $M_{\rm disc}$.
			Unfilled markers represent discs for which the mass inside the Keplerian
			disc differs from the numerical one by more than $0.2$. Massive discs have
			a shallower decline corresponding to smaller values of $\alpha's$.}
		\label{fig:alp_Mdisc}
	\end{figure}
	In \reffig{fig:alp_Mdisc}, we show the power-law exponent $\alpha$, obtained by
	fitting \refeq{eq:mass_J_keplerian} over the numerical data as a function of
	$M^{\rm num}_{\rm disc}$. Unfilled markers represent discs for which the mass of
	the Keplerian disc differs from the actual one by more than $0.2$. The exponent
	$\alpha$ changes considerably within our sample, from 4 up to 14, and more
	massive discs ($M_{\rm disc} > 10^{-2} \msun$) have a shallower decline,
	characterised by $4.0 \lesssim \alpha \lesssim 5.4$.
	\begin{figure}
		\includegraphics{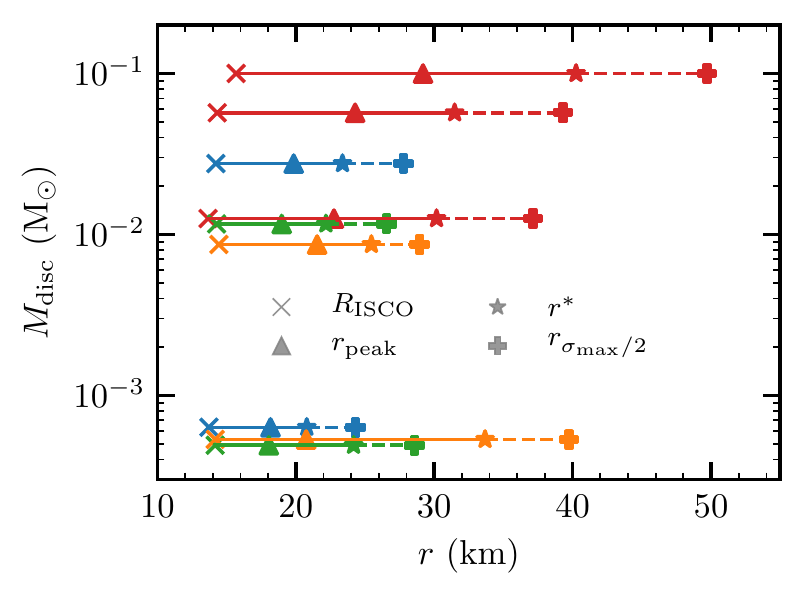}
		\caption{Fitted values of $R_{\rm ISCO}$, $r_{\rm peak}$ and $r^*$ as
			defined in \refeq{eq:disc_radial_density} for the discs reported in
			\ref{fig:keplerian_disc} except the simulation with error on the mass
			above 0.2. Solid lines represent the radius spanned by the Gaussian, while
			dashed lines represent the power decay branch of $\sigma(r)$ up to the
			radius $r_{\sigma_{\rm max}/2}$ at which the value of the density is half
			of its maximum.}
		\label{fig:params_Mdisc}
	\end{figure}
	The relevant parameters for the radial distributions of simulations at SR are
	summarised in \reffig{fig:params_Mdisc}. The radius of the ISCO $R_{\rm ISCO}$
	(crosses), of the density peak $r_{\rm peak}$ (up-triangles), of the junction
	between the Gaussian and the power decay $r^*$ (stars) and of the half density
	peak $r_{\sigma_{\rm max}/2}$ span a small range, indicating similar radial
	density distributions despite the mass spans almost 3 order of magnitude.
	$R_{\rm ISCO}$ is found at $13-16~{\rm km}$ from the centre, while the density
	peak is around $17-29~{\rm km}$.

	\section{Comparison with the fitting formulae used to compute GW190425 kilonova light curves}\label{sec:fitting_formulae}
	
	\begin{figure}
		\includegraphics{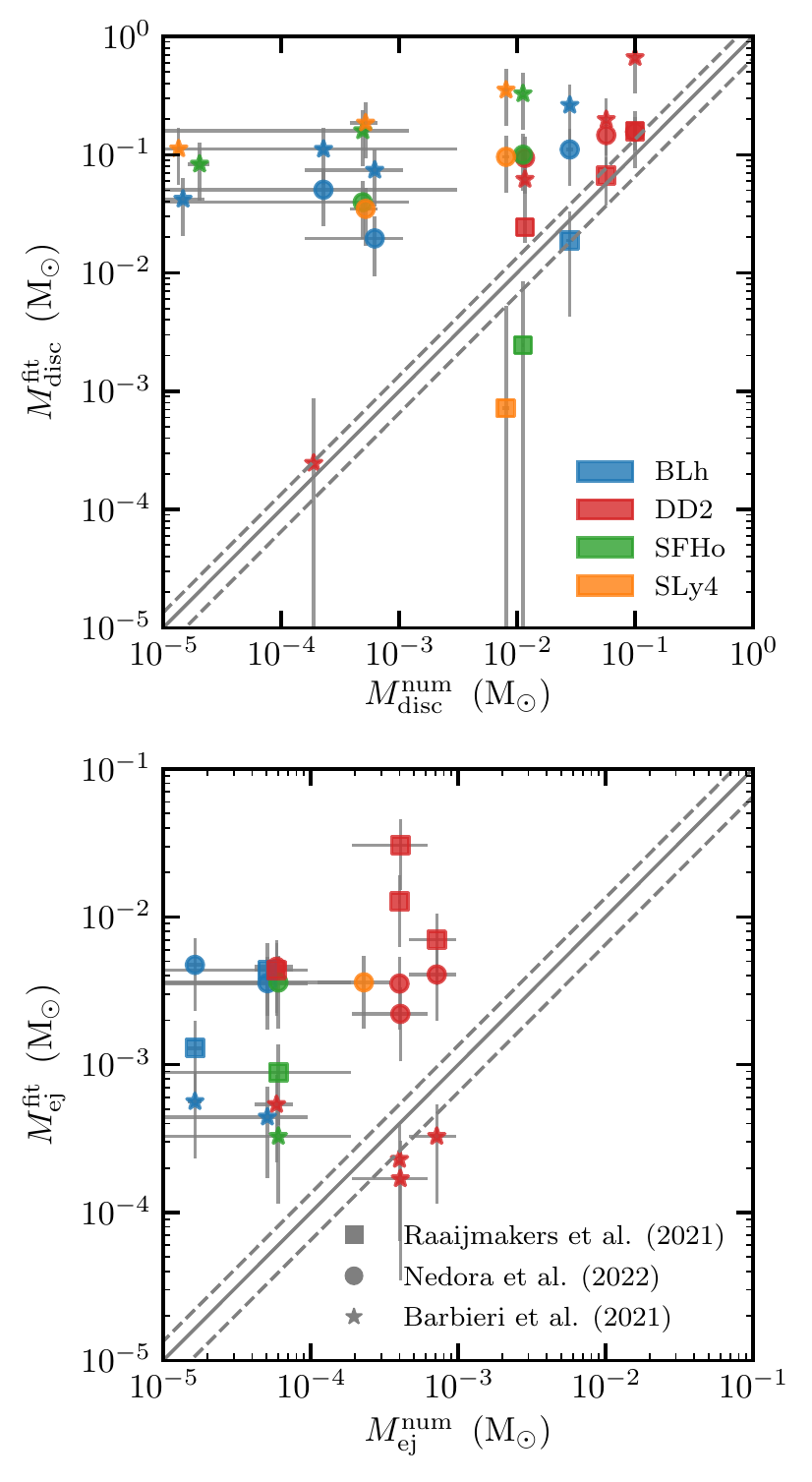}
		\caption{Top: Comparison of the disc masses obtained from our numerical simulations and from the fitting formulae used in \citet{Raaijmakers:2021slr} (originally, from \citet{Kruger:2020gig}) and in \citet{Barbieri.etal:2021}. Bottom: Comparison of the dynamical ejecta masses obtained from our numerical simulations and from the fitting formulae used in \citet{Raaijmakers:2021slr} (originally, from \citet{Kruger:2020gig}) and in \citet{Barbieri.etal:2021} (originally from \citet{Radice:2018pdn}). Fitting formulae from \citet{Nedora:2020qtd} are also reported. The error bars on the vertical (horizontal) axis are estimated as the 50 per cent of the predicted value (absolute difference between the SR and LR values). For the BNS in our sample with $M_{\rm disk}^{\rm num} \lesssim 10^{-3}~\Msun$ ($M_{\rm disk}^{\rm num} \lesssim 10^{-4}~\Msun$), the formulae from \citet{Kruger:2020gig} \citep{Nedora:2020qtd} result in nonphysical values for the disc mass.}
		\label{fig:NR_vs_predicted}
	\end{figure}
	
	In this appendix, we test the fitting formulae for the ejecta and disc properties used in \citet{Raaijmakers:2021slr} and \citet{Barbieri.etal:2021} in the parameter range of GW190425 to predict the associated kilonova light curves. Some of these formulae were originally proposed in \citet{Foucart:2016vxd}, \citet{Kruger:2020gig}, \citet{Radice:2018pdn} (see also \citet{Dietrich:2016fpt}). Additionally, we include in the comparison 
	fitting formulae from \citet{Nedora:2020qtd} in the form of their equation~6, i.e., a second-order
	polynomial in the mass ratio and tidal deformability. In particular, we use coefficients fitted on the dataset \texttt{RefM0Set \& M0/M1Set}, i.e., on a set of simulations
	including neutrino emission and absorption, and microphysical EOSs.
	We stress that we examine the different formulae in an unexplored parameter region since the binary systems within the calibration dataset are overall lighter and involve more deformable objects than those in our simulations.
	
	In \reffig{fig:NR_vs_predicted}, we compare the disc (top) and ejecta (bottom) masses predicted by the various fitting formulae with the ones obtained by our simulations. 
	The uncertainties in the fitted values are $50$ per cent of the estimated value, summed to a floor value of $5 \times 10^{-4} \Msun$ for the disc mass and $5 \times 10^{-5} \Msun$ for the ejecta mass. The bisector is the ``agreement line", while the dashed lines represent the $35$ per cent deviation from the exact prediction. For the mass of the dynamical ejecta only simulations with $\md > 10^{-5}\Msun$ have been taken into account. 

	In most of the cases, the fitting formulae significantly overestimate both the mass of the disc and the mass of the dynamical ejecta, and sometimes even predict opposite trends with respect to the binary parameters.
	Only in the case of the disc masses predicted by \citet{Kruger:2020gig} 
	\citep[used in][]{Raaijmakers:2021slr} and of the ejecta masses by \citet{Radice:2018pdn} 
	\citep[used in][]{Barbieri.etal:2021} there is a partial agreement, at least within the estimated uncertainties.
	
	The estimates of \citet{Nedora:2020qtd} is rather insensitive to the detailed binary parameters, giving rather similar ejecta mass and disc mass for each binary configuration.
	
	\begin{figure}
		\includegraphics{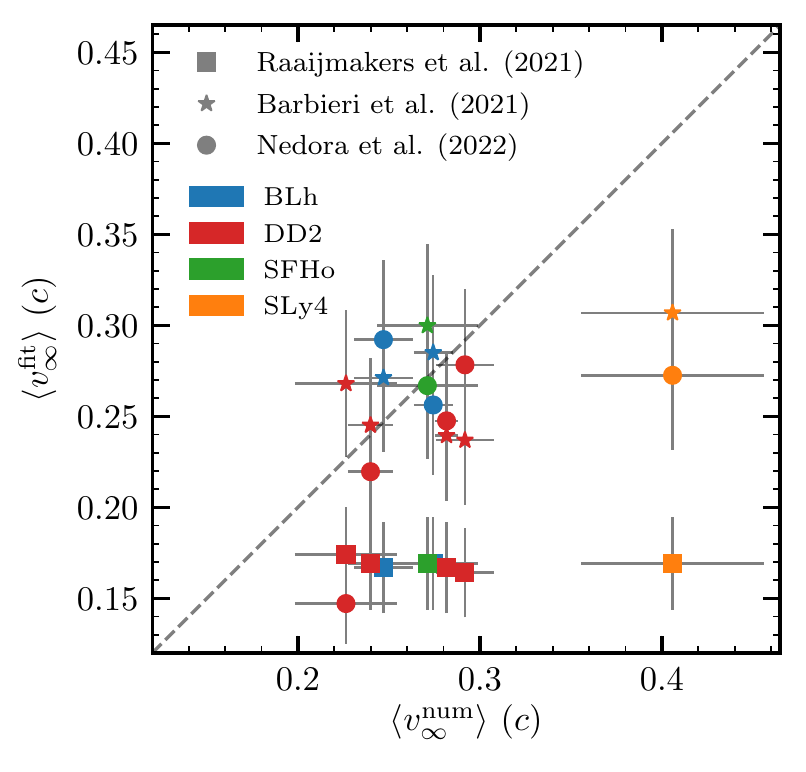}
		\caption{Comparison of the mass-weighted average velocity of the dynamical
			ejecta as obtained in our simulations and from the fitting formulae employed in the kilonova calculations of \citet{Raaijmakers:2021slr}
			and \citet{Barbieri.etal:2021}, taken from \citet{Foucart:2016vxd} and \citet{Radice:2018pdn}, respectively. Results from the fitting formulae from \citet{Nedora:2020qtd} are also reported.
			The (symmetric) uncertainties on the vertical axis are conservatively estimated as the 30 per cent of the values obtained from the fitting formulae.
			Error bars on the horizontal axis are estimated as the difference between the values inferred from the SR and LR simulations.}
		\label{fig:velocity_fit}
	\end{figure}
	
	Another physical input needed in kilonova light curves calculations is the
	velocity at which ejected matter is expelled from the binary system. In
	\reffig{fig:velocity_fit}, we show the mass-weighted average asymptotic velocity
	of the dynamical ejecta obtained from our numerical simulations and from the
	fitting formulae presented in \citet{Radice:2018pdn, Foucart:2016vxd,Nedora:2020qtd}. Only simulations with $\md > 10^{-5}\Msun$ have been taken into account.
	We assume a conservative uncertainty of the 30 per cent on the values obtained from the fitting formulae.
	We observe that the formulae from \citet{Radice:2018pdn} and \citet{Nedora:2020qtd} work reasonably well for outflow speed with $\langle v_\infty^{\rm num}\rangle$ in the range $0.24-0.30\,c$, while they
	underestimate the average velocity in the simulation with the fastest ejecta. The fitting
	formula from \citet{Foucart:2016vxd} used in \citet{Raaijmakers:2021slr} to make predictions on the kilonova from the GW190425 event, but originally tailored for the
	dynamical ejecta of BHNS systems, predicts a very similar average velocity for
	all the binaries, that is systematically smaller than the outcome of the
	simulations. This is because the expression assumes that the average velocity of
	the ejecta is given by a constant value of $\sim0.15$ plus a linear correction
	in the mass ratio, which is tiny in the case of BNS systems ($q\sim1-2$).

	\section{Standard deviation of the azimuthal angle}\label{sec:appendix_rms}
	The azimuthal angle of the dynamical ejecta distribution $\pd$ has a $2
	\pi$-rotational symmetry. So its mass weighted SD $\sdpd$ depends on an
	arbitrary chosen reference. For each angular bin $\phi_i$ of normalised weight
	$w_i$ of the ejecta distribution we define the periodic shift
	$S_{\delta}(\phi_i)$ as:
	\begin{equation}
	S_{\delta}(\phi_i) := 
	\begin{cases}
	\phi_i + \delta &{\rm if} \; \phi_i < 2\pi - \delta ~,\\
	\phi_i + \delta - 2\pi &{\rm if} \; \phi_i \geq 2\pi - \delta ~.\\
	\end{cases}
	\end{equation}
	Let's indicate with $S_{\delta}(\pd)$ the distribution obtained after the shift
	of awl the $\phi_i$.The average $\langle \pd \rangle_{\delta} \equiv \langle
	S_{\delta}(\pd) \rangle$ is then
	\begin{equation}
	\langle \pd \rangle_{\delta} = \langle \pd \rangle_0 + \delta - 2\pi W_{\delta} \, ,
	\end{equation}
	where $W_{\delta}$ is the total weight of the bins $\phi_i \geq 2\pi - \delta$,
	\begin{equation}
	W_\delta =  \sum_{\phi_i \geq 2\pi - \delta} w_i \leq 1 \, .
	\end{equation}
	We choose $\delta = \delta^*$ such that $\langle \pd \rangle_{\delta}$ is
	centred in the half of the interval, i.e in $\pi$~\footnote{Multiple $\delta^*$
		that satisfy this condition can exist, so we also add the condition that the
		mode of the distribution lies in the interval $\pi - \pi/4 \leq \phi \leq \pi +
		\pi/4$.}:
	\begin{equation}
	\delta^* - 2\pi W_{\delta^*} = \pi - \langle \pd \rangle_0\, .
	\end{equation}
	The root mean square (RMS) of $\pd$ after the shift $S_{\delta}$ is
	\begin{equation}
	\begin{split}
	{\rm RMS}_{\delta}(\pd) =\bigg[ &{\rm RMS}_0(\pd)^2 + 2 \delta \langle \pd \rangle_0 + \delta^2 + \\
	&+ 4\pi \left( (\pi - \delta) W_{\delta} - \overline{\langle \pd \rangle}_{\delta} \right)  \bigg]^{1/2} \, ,
	\end{split}
	\end{equation}
	where ${\rm RMS}_0(\pd)$ and $\langle \pd \rangle_0$ are the unshifted RMS and
	average of $\phi$ and $\overline{\langle \pd \rangle}_{\delta}$ is the average
	of the bins $\phi_i \geq 2\pi - \delta$,
	\begin{equation}
	\overline{\langle \pd \rangle}_{\delta} = \sum_{\phi_i \geq 2\pi - \delta} w_i \phi_i \, .
	\end{equation}
	Finally, the SD with respect to the new average $\langle \pd \rangle_{\delta}$
	is
	\begin{equation}
	\begin{split}
	\sigma_{\delta}(\pd) &= \sqrt{\sum_i w_i (S_{\delta}(\phi_i) - \langle \pd \rangle_{\delta})^2} \\
	& = \sqrt{{\rm RMS}_{\delta}(\pd)^2 - \langle \pd \rangle_{\delta}^2} \, .
	\end{split}
	\end{equation}

	\bsp    
	\label{lastpage}
\end{document}